\documentclass[]{aa}
\usepackage{graphicx}
\usepackage{amsmath,amssymb}
\usepackage{natbib}
\usepackage{txfonts}
\usepackage[colorlinks=true,linkcolor=blue,citecolor=blue]{hyperref}

\begin{document}

\title{Possible depletion of metals into dust grains in the core of the Centaurus cluster of galaxies}
\titlerunning{Dust depletion of metals in Centaurus}
   \author{K. Lakhchaura
          \inst{1,2}
          \and
          F. Mernier
          \inst{1,3}
          \and
          N. Werner
          \inst{1,4,5}
          }

   \institute{MTA-E\"otv\"os University Lend\"ulet Hot Universe Research Group, P\'azm\'any P\'eter s\'et\'any 1/A, Budapest, 1117, Hungary\\
              \email{lakhchaura.k@gmail.com}
         \and
             MTA-ELTE Astrophysics Research Group, P\'azm\'any P\'eter s\'et\'any 1/A, Budapest, 1117, Hungary
        \and
            SRON Netherlands Institute for Space Research, Sorbonnelaan 2, 3584 CA Utrecht, The Netherlands
        \and
            Department of Theoretical Physics and Astrophysics, Faculty of Science, Masaryk University, Kotl\'a\v{r}sk\'a 2, Brno, 611 37, Czech Republic
        \and
            School of Science, Hiroshima University, 1-3-1 Kagamiyama, Higashi-Hiroshima 739-8526, Japan
             }

   \date{Received 30 November 2018 / Accepted 12 December 2018}


\abstract{
We present azimuthally averaged metal abundance profiles from a full, comprehensive, and conservative re-analysis of the deep ($\sim$800 ks total net exposure) \textit{Chandra}/ACIS-S observation of the Centaurus cluster core (NGC\,4696). After carefully checking various sources of systematic uncertainties, including the choice of the spectral deprojection method, assumptions about the temperature structure of the gas, and uncertainties in the continuum modeling, we confirm the existence of a central drop in the abundances of the `reactive' elements Fe, Si, S, Mg, and Ca, within $r\lesssim$10 kpc. The same drops are also found when analyzing the \textit{XMM-Newton}/EPIC data ($\sim$150 ks). Adopting our most conservative approach, we find that, unlike the central drops seen for Fe, Si, S, Mg and Ca, the abundance of the `nonreactive' element Ar is fully consistent with showing no central drop. This is further confirmed by the significant ($>3\sigma$) central radial increase of the Ar/Fe ratio. Our results corroborate the previously proposed `dust depletion scenario' , in which central metal abundance drops are explained by the deposition of a significant fraction of centrally cooled reactive metals into dust grains present in the central regions of the Centaurus cluster. This is also supported by the previous findings that the extent of the metal abundance drops in NGC\,4696 broadly coincides with the infrared dust emission.
}


\maketitle

\section{Introduction} 
\label{sec:intro}

One of the most compelling ways to study the chemical evolution of the universe at its largest scales is via X-ray spectroscopy of galaxy clusters, groups, and giant elliptical galaxies. Indeed, the gravitational potential wells of these systems are deep enough to retain the major fraction of their baryons, which consist of a hot ($10^6$--$10^8$ K), tenuous ($10^{-1}$--$10^{-5}$ cm$^{-3}$) X-ray-emitting plasma \citep[for a review, see e.g.,][]{Boehringer2010}. These hot atmospheres are rich in heavy elements (C, N, O, Ne, Mg, Si, S, Ar, Ca, Cr, Mn, Fe, and Ni) whose abundances can be relatively well constrained from their K-shell emission lines in the X-ray energy window \citep[for a recent review, see][]{Mernier2018}. 

Thanks to current X-ray observatories like \textit{Chandra} and \textit{XMM-Newton,}  which allow to do accurate spatial spectroscopy, the spatial distribution of metals in the diffuse intracluster medium (ICM) can be studied in detail, providing invaluable information on where, how, and when these metals -- produced by Type Ia (SNIa) and core-collapse supernovae (SNcc) -- escaped from their galaxy hosts and enriched hot atmospheres of galaxy clusters, groups, and ellipticals. In the outskirts, the recent discovery of flat, uniform metal abundance profiles \citep{Werner2013b,Simionescu2015,Urban2017} strongly suggests that the bulk of metals enriched the hot ICM before clusters started to form and attain a well stratified entropy  (i.e., typically at $z \gtrsim 2$-$3$). This scenario is also supported by recent cosmological hydrodynamical simulations \citep{Biffi2017,Biffi2018a,Truong2018}, suggesting that active galactic nuclei (AGNs) play an essential role in spreading metals out of their host galaxies before assembling into clusters \citep[for a review, see e.g.,][]{Biffi2018b}. In addition, it is now well established that cool-core clusters\footnote{The term `cool-core' is used for systems in which the cooling time of the central ICM is typically shorter than the Hubble time \citep{Molendi2001b,Hudson2010}.} host a centrally peaked Fe distribution while noncool-core clusters exhibit a much flatter Fe profile \citep{DeGrandi2004,Lovisari2018}. The central Fe excess found in cool-core clusters was initially thought to be produced by ongoing SNIa explosions from the low-mass stellar population within the (red-and-dead) brightest cluster galaxies \citep[BCG; e.g.,][]{Boerhinger2004}. However, the surprising finding that SNcc products (i.e., O, Mg, as well as a fraction of Si, S, and Ar), which are not expected to be produced by BCGs within the last $\sim$7-8 Gyr, also exhibit a central peak reopened the question of the origin of the central enrichment in cool-core systems \citep{Simionescu2009,Mernier2017}.

\vspace{0.7cm}

Perhaps even more intriguing is the presence of apparent metal abundance drops in the very core (i.e., within a few kiloparsecs) of some relaxed systems. First discovered in the galaxy cluster Abell\,2199 \citep{Johnstone2002}, central drops were later reported for Fe in many other sources, including also less massive groups and ellipticals \citep[e.g.,][]{Churazov2003,Churazov2004,Rasmussen2007,Rafferty2013,Panagoulia2015,Mernier2017,Gendron-Marsolais2017}. Because these drops are difficult to explain with simple enrichment models, they were initially thought to be artefacts of spectral analysis. For instance, \citet{Werner2006} showed that the apparent drop seen in 2A\,0335+096 was solely due to the oversimplified assumption of a single-temperature plasma, as no inversion in the Fe profile is found when two or more temperatures are modeled in the spectra \citep[i.e., the so-called ``Fe-bias''; e.g.,][]{Buote2000}. In several other cases, however, assuming a multi-temperature structure for the gas does not help to attenuate these drops. Possible effects from resonance scattering have also been proposed, but were found to contribute to only a negligible fraction of the apparently missing Fe \citep{Sanders2006a,Gendron-Marsolais2017}. Besides Fe, central abundance drops were also found for several other elements, in particular Si and S \citep{Panagoulia2015,Mernier2017}.

Under the assumption that the observed metal abundance drops are real, it has been proposed \citep{Panagoulia2013,Panagoulia2015} that a significant fraction of metals initially residing in the hot phase will rapidly cool down and be deposited into dust grains, thereby becoming invisible in the X-ray band. As most BCGs host an AGN which is able to uplift and re-heat its surrounding hot atmosphere via mechanical feedback \citep[for a recent review, see][]{Werner2018b}, metal-rich dust grains may be dragged out of the innermost regions, possibly in the form of extended emission line filaments, before becoming re-heated and returning into their initial hot phase a few kiloparsecs away from the center.

A key prediction of the above scenario is that, unlike all the metal abundances that are measurable in X-ray (i.e., `reactive' elements), abundance drops are not expected to occur for Ne and Ar, as these two elements are noble gases and are not easily deposited into dust grains (i.e., `non-reactive'). The current CCD instruments do not have the required spectral resolution to  accurately constrain the Ne abundance\footnote{Unfortunately, the main Ne emission lines (\ion{Ne}{x} and \ion{Ne}{xi}) are located within the Fe-L complex unresolved by CCD instruments, making the Ne abundance strongly degenerate with other parameters, e.g., the Fe abundance or the temperature structure of the gas. While the Reflection Grating Spectrometer (RGS) instrument on-board \textit{XMM-Newton} is technically able to resolve these lines, its slitless design does not allow for spatially resolved spectroscopy.} and no firm conclusion could be drawn from the central Ar profiles using \textit{XMM-Newton} observations of nearby cool-core systems \citep{Mernier2017}. In addition to the moderate spatial resolution of \textit{XMM-Newton}, the main reason is that stacking data of many systems naturally increases the intrinsic scatter of the profiles, especially in cool-core systems \citep[see also][]{Lovisari2018}. Clearly, testing the above scenario requires very deep observations of one bright, carefully selected, nearby cool-core system.

The most striking example of central metal abundance drops certainly comes from Abell\,3526 (i.e., the Centaurus cluster). The central BCG (NGC\,4696) of this bright and nearby cool-core system exhibits one of the most prominent Fe drops known to date (with an abrupt inwards decrease from $\sim$2 to $\sim$0.5 solar), which was the second ever discovered \citep{Sanders2002}. Since then, additional detailed observations of the Centaurus cluster have further confirmed this remarkable drop within its central 5-20 kpc, not only in Fe but also in Si, S, and possibly Mg \citep{Sanders2006b,Panagoulia2013,Sanders2016}. In addition, NGC\,4696 is well known to contain significant amounts of dust, observed in far-infrared with \textit{Spitzer} \citep{Kaneda2005,Kaneda2007} and  \textit{Herschel} \citep{Mittal2011}, as well as extended H$\alpha$ and optical (in particular [NII]) filaments \citep[][see also Lakhchaura et al. \citeyear{Lakhchaura2018}]{Fabian2016,Hamer2018}. 

In this paper, we take advantage of the extremely deep \textit{Chandra} observations of NGC\,4696 (residing in the core of the Centaurus cluster), obtained in successive pointings over 14 years since the beginning of the mission, combined with the unique spatial resolution of the telescope, in order to infer unprecedented constraints on its metal abundance profiles. Specifically, we aim to compare the (non-reactive) Ar radial distribution with that of the other (reactive) elements in order to test the dust depletion scenario as a possible explanation of the central metal abundance drops. In this respect, we put a strong emphasis on the major systematic effects that may alter such measurements and possibly explain some results previously reported in the literature. This includes a careful comparison with \textit{XMM-Newton} measurements, which we also re-analyze.
The data reduction and analysis of all the \textit{Chandra} and \textit{XMM-Newton} observations are described in Sect. \ref{sec:data}. Our results are reported in Sect. \ref{sec:results}, and are subsequently discussed and further interpreted in Sect. \ref{sec:discussion}. We summarize our conclusions in Sect. \ref{sec:conclusion}. Throughout this paper, we adopt a $\Lambda$CDM cosmology with H$_{0}=$ 70 km s$^{-1}$ Mpc$^{-1}$, $\Omega_{\rm M}=$ 0.3, and $\Omega_{\Lambda}=$ 0.7. The metal abundances are given with respect to the solar values of \citet{Grevesse1998}. Unless stated otherwise, the errors are given within their 68.27\% confidence level (1$\sigma$).

\section{Data}
\label{sec:data}

For this work, we used all the publicly available \textit{Chandra} and \textit{XMM-Newton} observations of the core of the Centaurus cluster (NGC\,4696), taken from the High-Energy Astrophysics Science
Archive Research Center (HEASARC). A log of these observations is given in Table~\ref{tab:obsn_log}.

\begin{table}
 \caption{A log of the \textit{Chandra} and \textit{XMM-Newton} observations of NGC\,4696, used in this paper.}
\label{tab:obsn_log}
\centering
{\footnotesize
\begin{tabular}{c c c c c}
\hline
Obs ID & Instrument & Cleaned Exp. & Date of \\
 & & (ks) & Observation\\
\hline
\multicolumn{4}{c}{\textit{Chandra}} \\
\hline
504 & ACIS-S & 31.48 & 2000-05-22\\
505 & ACIS-S &  9.96 & 2000-06-08\\
1560 & ACIS-S & 45.56 & 2001-04-18\\
4190 & ACIS-S & 34.00 & 2003-04-18\\
4191 & ACIS-S & 32.74 & 2003-04-18\\
4954 & ACIS-S & 87.26 & 2004-04-01\\
4955 & ACIS-S & 44.68 & 2004-04-02\\
5310 & ACIS-S & 36.24 & 2004-04-04\\
16223 & ACIS-S & 175.38 & 2014-05-26\\
16224 & ACIS-S & 40.76 & 2014-04-09\\
16225 & ACIS-S & 29.33 & 2014-04-26\\
16534 & ACIS-S & 54.68 & 2014-06-05\\
16607 & ACIS-S & 44.78 & 2014-04-12\\
16608 & ACIS-S & 33.35 & 2014-04-07\\
16609 & ACIS-S & 80.54 & 2014-05-04\\
16610 & ACIS-S & 16.57 & 2014-04-27\\
\hline
\multicolumn{4}{c}{\textit{XMM-Newton}} \\
\hline
0046340101 & EPIC MOS\,1 & 44.22 & 2002-01-03\\
0046340101 & EPIC MOS\,2 & 44.32 & 2002-01-03\\
0046340101 & EPIC pn & 48.56 & 2002-01-03\\
0406200101 & EPIC MOS\,1 & 107.64 & 2006-07-24\\
0406200101 & EPIC MOS\,2 & 108.81 & 2006-07-24\\
0406200101 & EPIC pn & 92.61 & 2006-07-24\\
\hline
\end{tabular}}
\end{table}

\subsection{\textit{Chandra} data reduction and analysis}
\label{sec:xray_data_red_analys}

We used the Chandra Interactive Analysis of Observations (CIAO) software version 4.9 \citep{Fruscione2006} 
and CALDB version 4.7.3 for all the data reduction, and for the spectral analyses, we used the X-ray spectral fitting package XSPEC version 12.9.1 \citep{Arnaud1996}. The procedures used for the data reduction were the same as described in \citet{Lakhchaura2018}. 
The standard \texttt{chandra\_repro} tool was used to reprocess the data 
and strong background flares were filtered and removed using the \texttt{lc\_clean} script. Once filtered, the cleaned ACIS-S data have a total exposure of 797.31 ks (Table~\ref{tab:obsn_log}). The point sources in the field were detected using
the CIAO task \texttt{wavdetect} (false-positive probability threshold$=$10$^{-6}$), verified by visual inspection of the 
X-ray images and removed from the subsequent analysis.

\subsubsection{Spectral extraction}
\label{sec:spectral_extr}

\begin{figure*}
\centering
  \includegraphics[width=.8\linewidth,trim={0 0cm 0 2.5cm},clip]{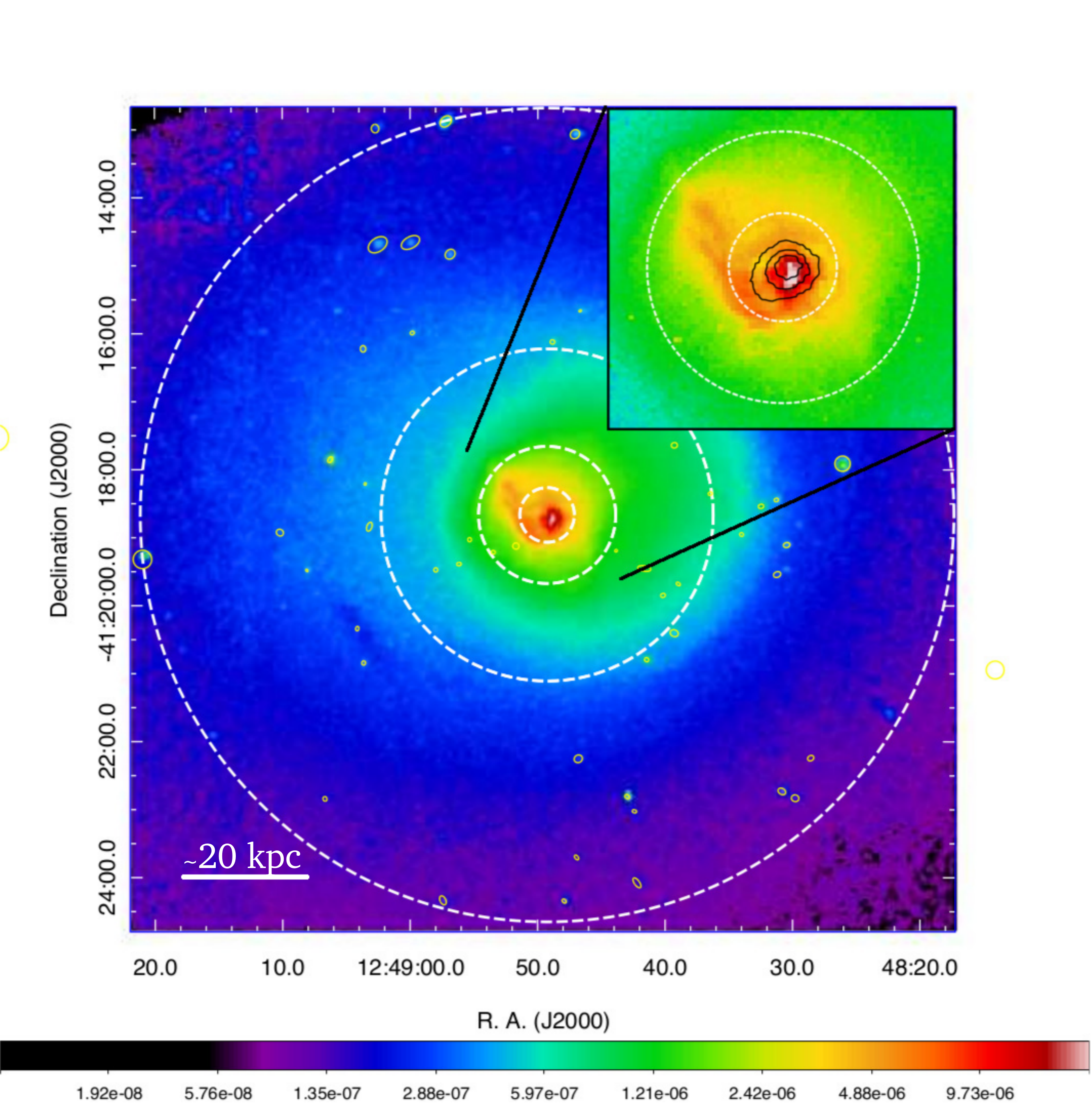}
\caption{An exposure corrected 0.5-7.0-keV \textit{Chandra} ACIS-S image of the Centaurus cluster. The color bar indicates the photon flux in counts/cm$^2$/s. The white dashed circles show the four annular regions used for the spectral extraction. The inset is a zoomed-in section of the central region showing the central soft X-ray filaments and the asymmetric distribution of the central X-ray emission. The black contours in the inset are from a {\it Herschel} PACS 100 $\mu m$ photometric observation of the cluster, and trace the central dust emission.}
\label{fig:centaurus_rgb}
\end{figure*}

Although the remarkable amount of data available for NGC\,4696 allows for the spectral regions to be 
subdivided into multiple, azimuthally resolved spatial bins \citep[e.g.,][]{Sanders2016}, here we aim to optimize the number of counts available per region in order to provide the most accurate constraints or the Ar abundance. This implies adopting radial bins that are both azimuthally averaged (i.e., concentric annuli instead of sectors) and much larger than what is usually considered based on the spatial resolution of \textit{Chandra}. 
Therefore, spectra were extracted from four concentric annuli (or shells, when the spectra are further deprojected), bounded by 4.4, 11.0, 26.6, and 65.3 kpc (corresponding to $\sim$0.4, 1.0, 2.5, 6.0 arcmin, respectively; see Fig.~\ref{fig:centaurus_rgb}) and centered on the X-ray emission peak, 
using the CIAO task \texttt{specextract}. The annuli selection lead to $\sim$0.76, 1.51, 2.61 and 3.71 million counts in the four annuli, respectively, in the 0.5--7.0 keV energy range.

    The standard \textit{Chandra} blank-sky background event files matching the source observations were
used to extract the background spectrum corresponding to each of the source spectra. The blank-sky event files were reprojected to match 
the source observation coordinate frame. We also scaled all the blank-sky spectra by the ratio of the 9.5-12 keV count rates of the source 
and blank-sky observations, in order to match the time-dependent particle background levels in the source and 
blank-sky observations. The differences in the Galactic foreground level in the scaled blank-sky and source spectra, determined using the 
{\it ROSAT} All Sky Survey 0.47--1.21 keV (RASS 45 band) count rates, were found to be insignificant. We note that, whereas background issues can sometimes be appreciable when analyzing extended X-ray sources, the exceptional brightness of NGC\,4696 makes background uncertainties largely negligible in the present analysis.

\subsubsection{Spectral fitting}
\label{sec:spectral_fit}
  The gas in the central regions of Centaurus is known to be multiphase. Therefore, single-temperature 
plasma models cannot fit the spectra well enough in the innermost bins, and may even cause significant underestimation of Fe due to ``Fe-bias'' \citep[][see also Mernier et al. \citeyear{Mernier2018}]{Buote2000}. To account for the complex multi-temperature structure of the cluster core, we successively fitted the spectra with two different multi-temperature models, namely a two-component \textit{vapec} model  (using AtomDB version 3.0.7), hereafter `2T \textit{vapec}', and  a 
\textit{vgadem} model, which reproduces a normally distributed differential emission measure. We also added an additional thermal 
bremsstrahlung component with the temperature fixed at 7.3 keV to account for the unresolved low-mass X-ray binaries in NGC\,4696 
\citep[see][]{Irwin2003}. The Galactic absorption was accounted for by using the photoelectric absorption model 
\textit{wabs}. The neutral hydrogen column density and the redshift were kept as free parameters in all the analyses. We verified that, within a given annulus, a slight variation of these two best-fit parameters has negligible effects (i.e., a few percent at most) on our results.

  For both the 2T \textit{vapec} and \textit{vgadem} models, the Fe, Si, S, Mg, Ca and Ar abundances were kept as free parameters, the He abundance was fixed to the solar value, and all other elements were tied to Fe. For 
the 2T \textit{vapec} model, the temperatures and normalizations of the two \textit{vapec} components were kept as free parameters while the abundances of the same elements were tied between the two components. We also  note that a single-temperature component (\textit{vapec}) was used when modeling two components did not significantly improve the fits (which often results in highly degenerate parameters between the two components). For the 
\textit{vgadem} model, the mean temperature, the width of the temperature distribution, and the normalization were kept as free parameters. 
To determine the gas density ($n=n_e+n_i$) of the four shells (assuming spherical symmetry), a constant density was assumed in 
each shell, and then the densities were obtained from the \textit{vapec}/\textit{vgadem} normalizations ($\eta$) using the following relation. 
\begin{equation}
\eta = \frac{10^{-14} \int{n_e n_p dV} }{4 \pi {D_{A}}^{2} (1+{\rm z})^2},
\end{equation}
where $D_A$ is the angular diameter distance of the source, and $n_e$ and $n_p$ are the electron and proton number densities 
(for a fully ionized gas with one third solar elemental abundance $n_e=0.53n$ and $n_p=n_e/1.2$), respectively.

\subsection{\textit{XMM-Newton} data reduction and analysis}
\label{sec:XMM_data_red}

We also compare the results obtained using the deep \textit{Chandra} data with the archival \textit{XMM-Newton} observations of the Centaurus core, which were taken in two separate pointings (January 2002 and July 2006). The data reduction was performed using the \textit{XMM-Newton} Science Analysis System (SAS) v14.0.0 and follows the same procedure as presented in \citet{Mernier2015}. In summary, we process the EPIC MOS (i.e., MOS\,1 and MOS\,2) and pn data using the tasks \texttt{emproc} and \texttt{epproc}, respectively, before filtering them to discard flared events. To do so, the light curves in the 10--12 keV band were stacked in 100-s binned histograms. We then fitted a Gaussian to the distribution and selected the count-rate threshold above which the events are rejected to be $\mu + 2\,\sigma$. The same procedure was repeated for the 0.3-2 keV band (for 10-s binned histograms), as flaring events may affect softer energies differently \citep{Lumb2002}. A summary of the cleaned \textit{XMM-Newton} data ($\sim$150 ks for the two pointings) is also presented in Table~\ref{tab:obsn_log}.

The rest of the analysis of the \textit{XMM-Newton} data follows that described above for \textit{Chandra} (Sects. \ref{sec:spectral_extr} and \ref{sec:spectral_fit}). Specifically, the EPIC spectra were extracted following the same regions as shown in Fig.~\ref{fig:centaurus_rgb} and the excluded point sources were selected from the ACIS-S detections. In order to account for the larger point spread function (PSF) of \textit{XMM-Newton}, the circular regions centered on the point sources were excised within a radius of 10 arcsec \citep{Mernier2015}. For the two \textit{XMM-Newton} observations, the combined 0.5-7.0 keV counts in the four annuli were $\sim$0.11, 0.25, 0.45 and 0.73 million for MOS\,1;  $\sim$0.11, 0.24, 0.46 and 0.76 million for MOS\,2; and $\sim$0.19, 0.36, 0.65 and 1.00 million for pn, respectively. The RMFs and ARFs were obtained via the SAS tasks \texttt{rmfgen} and \texttt{arfgen}, respectively. The MOS\,1, MOS\,2, and pn spectra were all fitted simultaneously, with all their parameters tied accordingly between the three instruments.

\section{Results}
\label{sec:results}

As stated in Sect. \ref{sec:intro}, we aim to carry out a comprehensive study on the central metal abundance drops in the Centaurus cluster, taking into account all the potential fitting biases that might significantly affect our measurements. In the following subsections, we compare our radial profiles obtained with different assumptions, namely the temperature structure of the gas, the choice of the spectral deprojection method,  the energy band within which the spectra are fitted, and the instruments used.

\subsection{Projected profiles: `vgadem' versus `2T vapec' models}
\label{sec:proj_prof}

First, we focus on the \textit{Chandra} observations of NGC\,4696 as directly seen by the ACIS-S instrument, that is, projected on the plane of the sky. As stated in Sect. \ref{sec:spectral_fit}, two types of multi-temperature models are considered: \textit{vgadem} and 2T \textit{vapec}. In the case of the 2T \textit{vapec} model, a single-temperature (\textit{vapec}) component was enough to fit the spectra beyond the two innermost annuli. The projected profiles of temperature (i.e., the mean temperature for \textit{vgadem} and the higher temperature for 2T \textit{vapec}), 
total number density, Fe, Si, S, Mg, Ca, and Ar abundances, and the Ar/Fe ratio, obtained using these two models successively, are shown in Table~\ref{tab:results} and Fig.~\ref{fig:deproj_vgadem_vs_2Tvapec} (red and blue filled areas).

Globally, these profiles seem to be very similar, and do not depend strongly on the choice of the multi-temperature model. We note, however, that the density is systematically higher in the 2T \textit{vapec} (by typically $\sim$30\%) than in the \textit{vgadem} case. As expected, it clearly appears that, no matter which model is considered, the Fe, Si, S and Mg profiles show a central decrease at $r\lesssim$10 kpc. Drops are also reported for Ar and Ca, although only at $r\lesssim$4~kpc. However, the Ar/Fe ratio (Fig.~\ref{fig:deproj_vgadem_vs_2Tvapec}, bottom right panel) seems to be 
increasing towards the center and the increase in the innermost bin, as compared to the outermost bin, is $>3\sigma$ and $>2\sigma$ when using the 2T \textit{vapec} and the \textit{vgadem} models, respectively.

\subsection{Deprojected profiles}
\label{sec:deproj_prof}

\begin{figure*}
\centering
  \includegraphics[width=.32\linewidth]{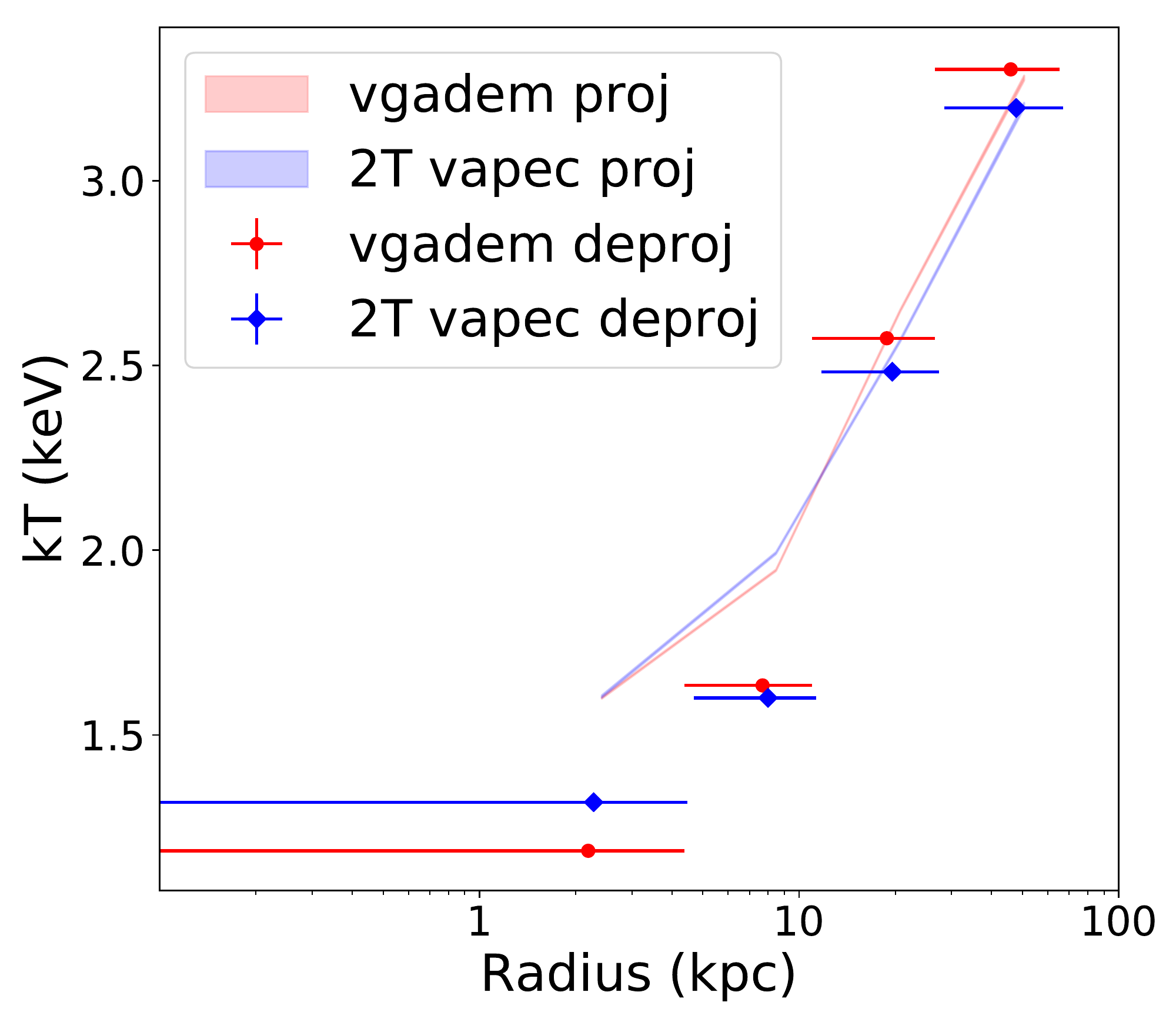}
  \includegraphics[width=.32\linewidth]{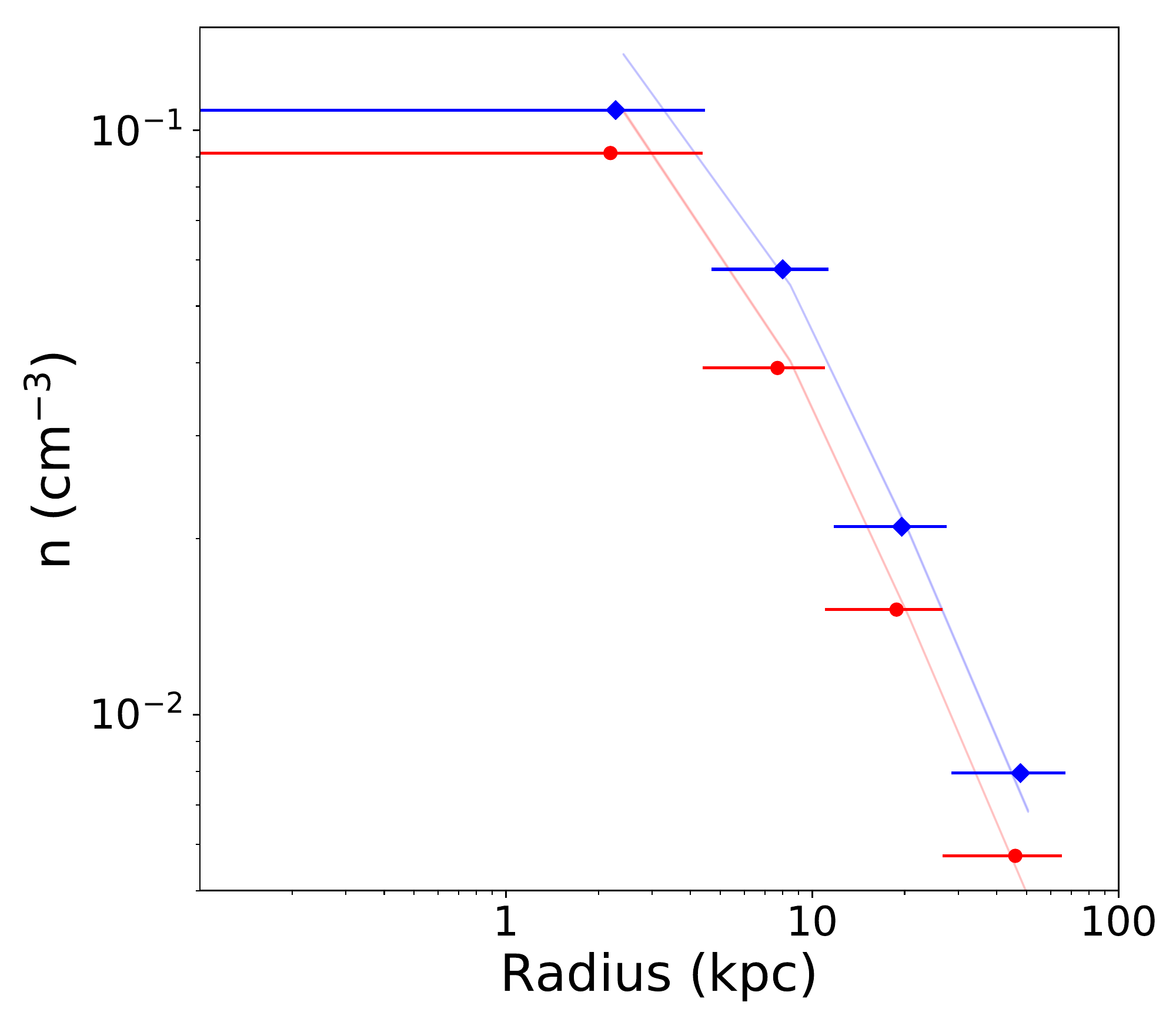}
  \includegraphics[width=.32\linewidth]{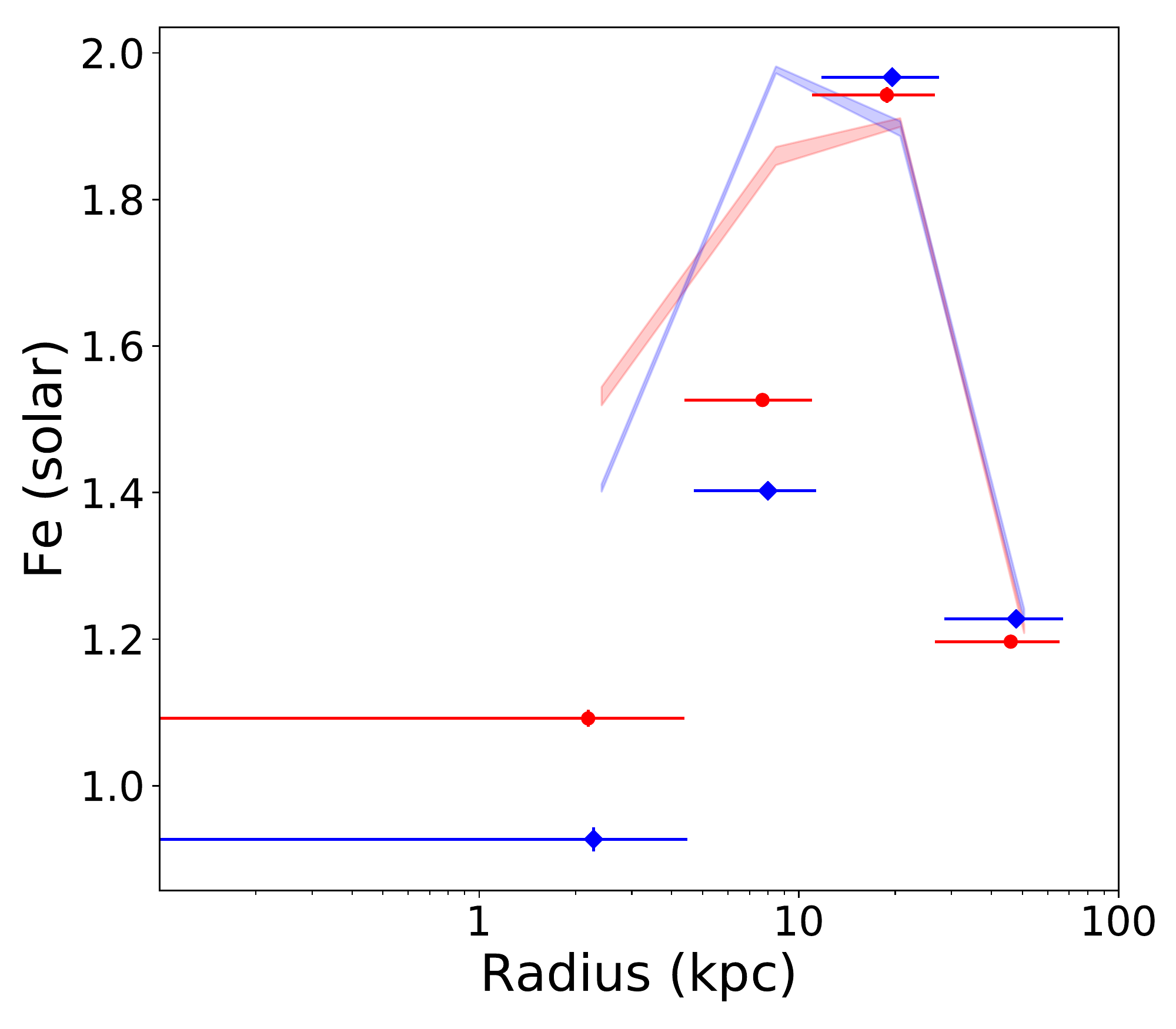}
  \includegraphics[width=.32\linewidth]{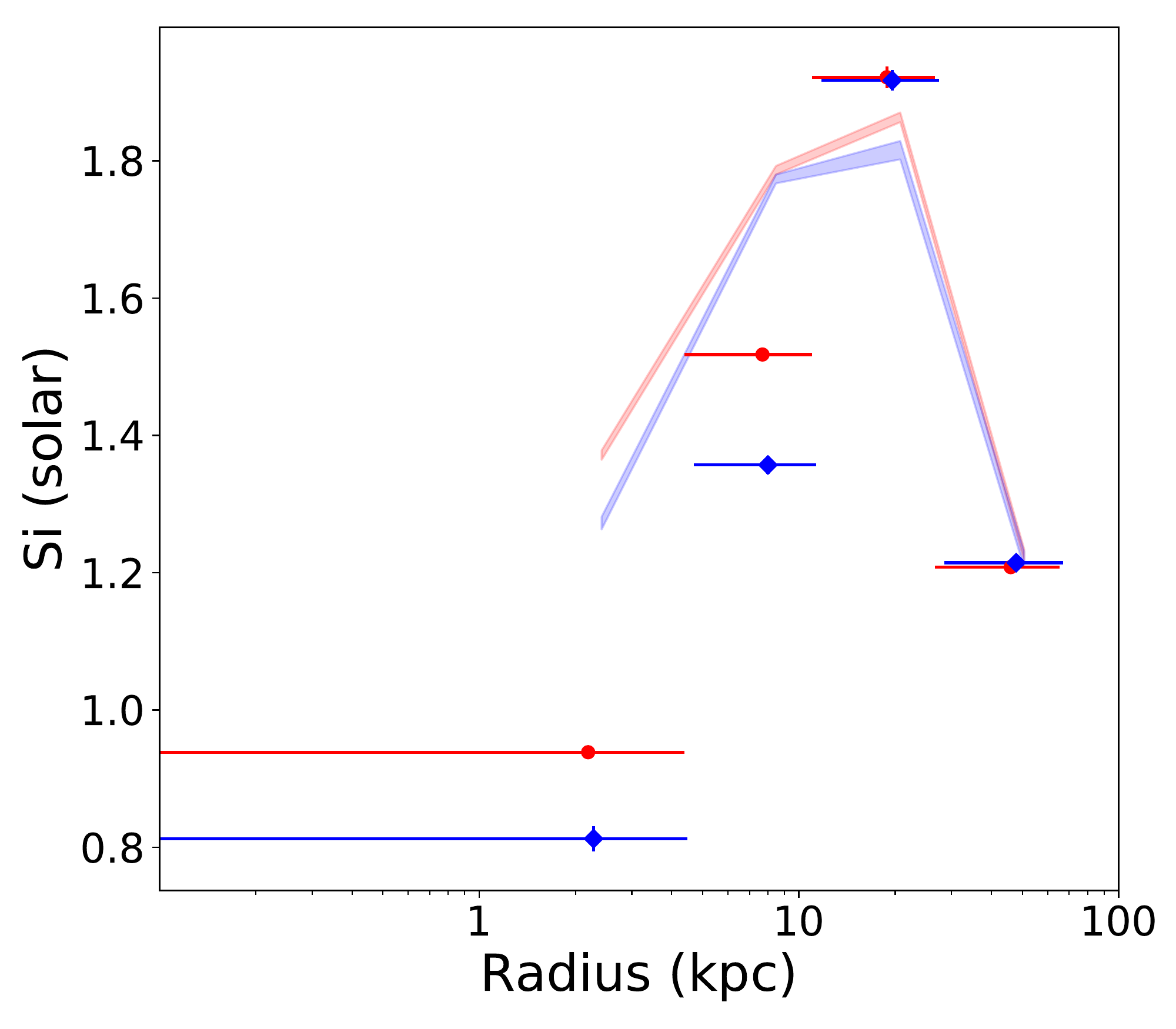}
  \includegraphics[width=.32\linewidth]{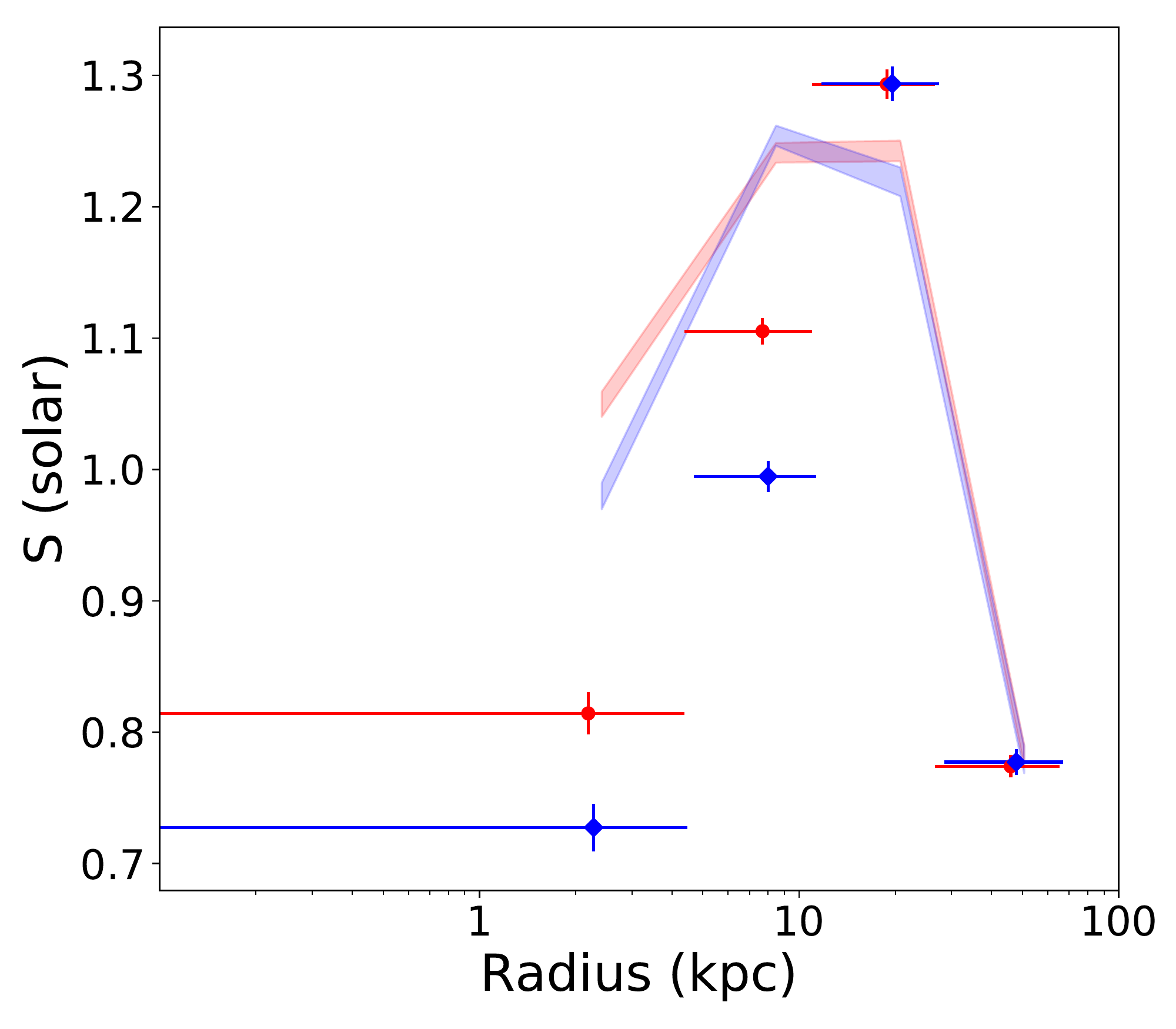}
  \includegraphics[width=.32\linewidth]{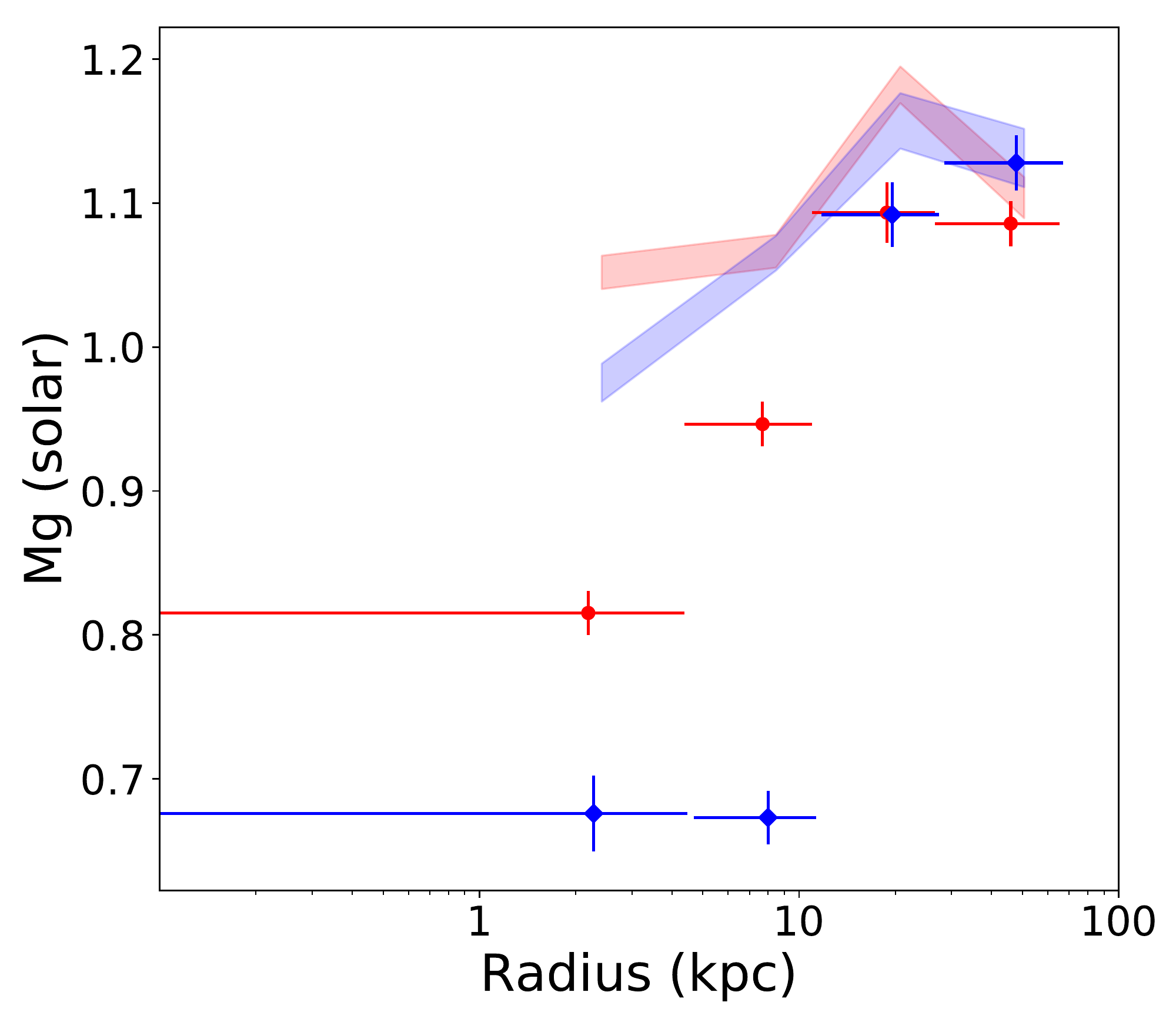}
  \includegraphics[width=.32\linewidth]{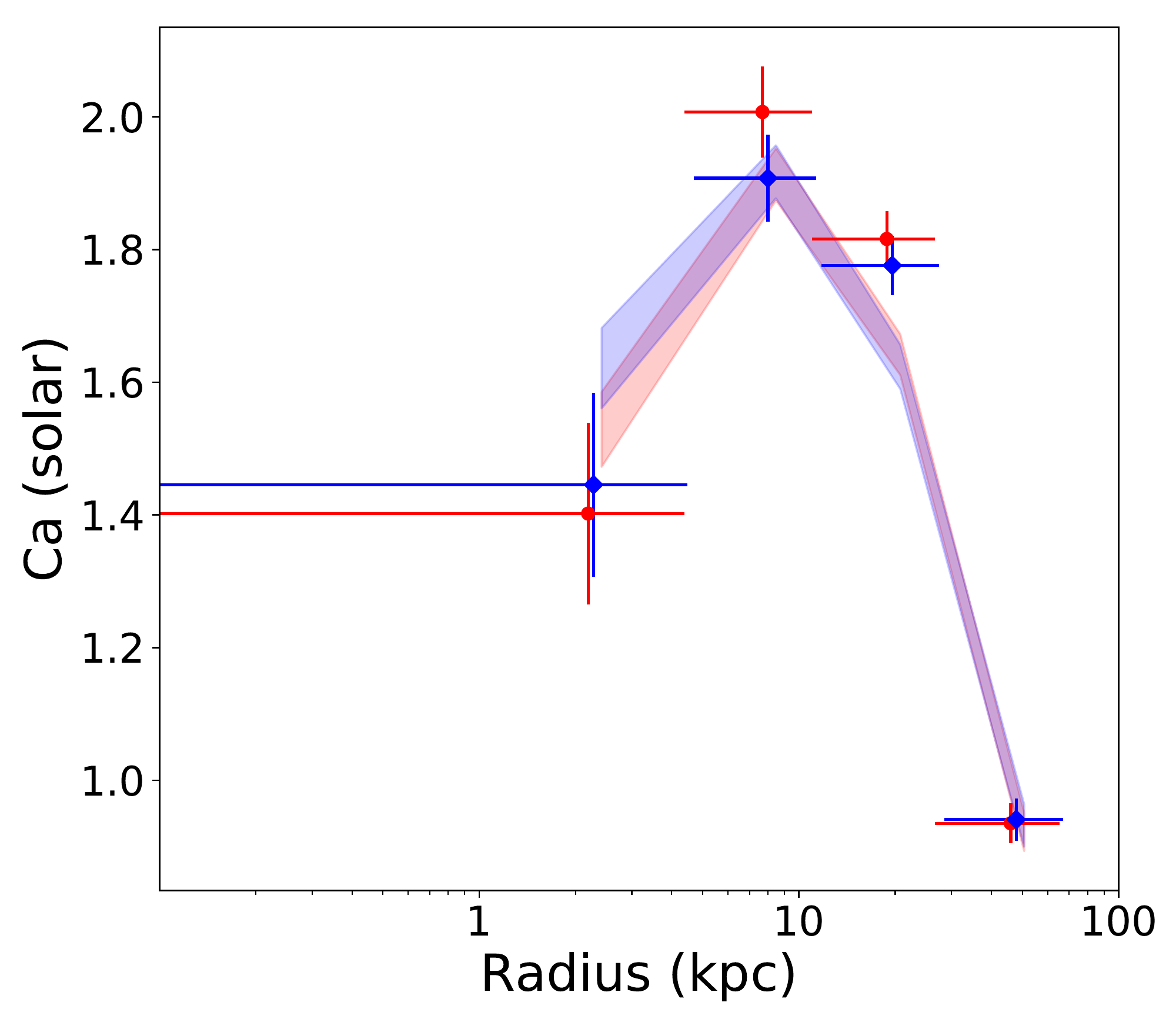}
  \includegraphics[width=.32\linewidth]{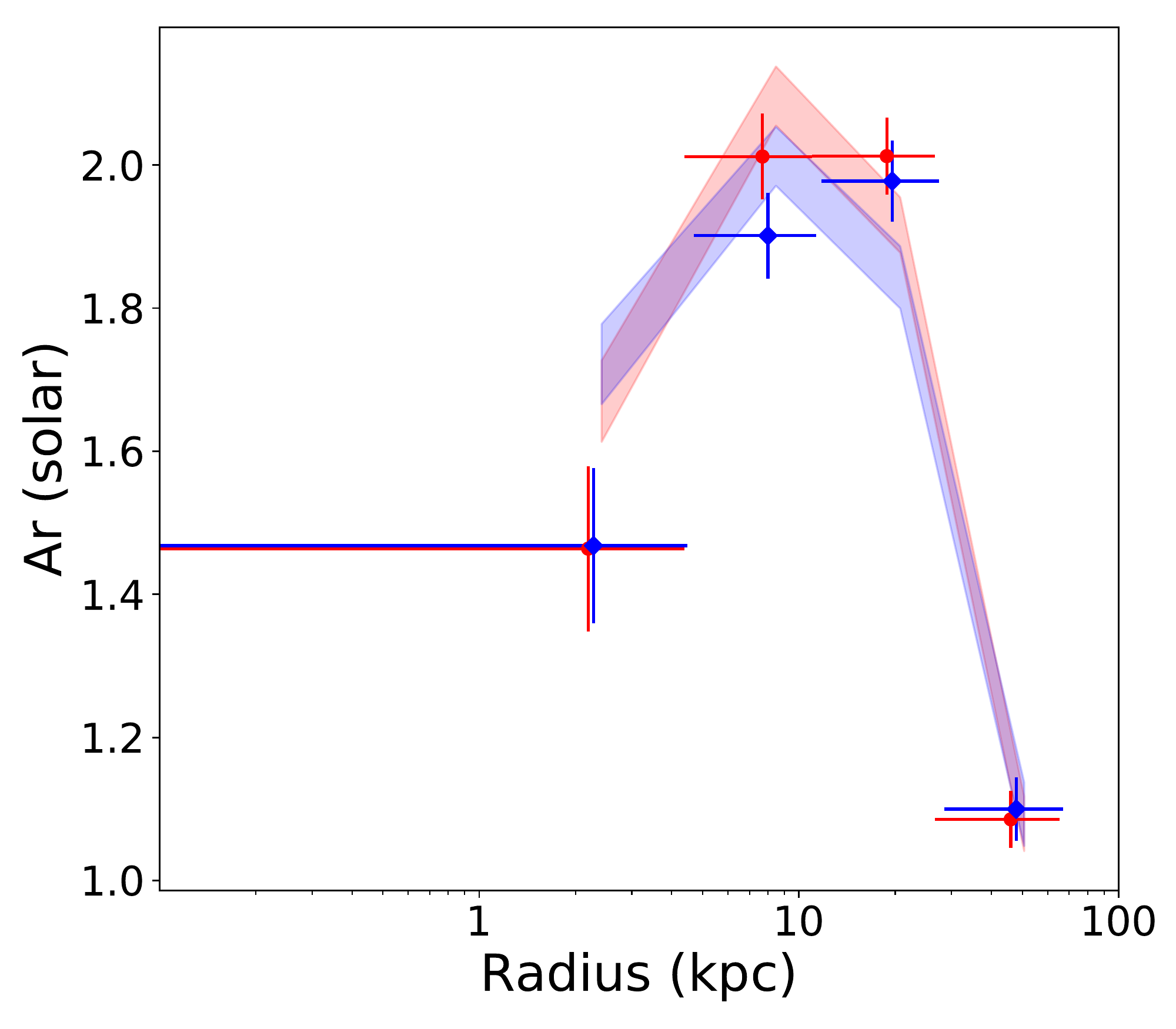}
  \includegraphics[width=.32\linewidth]{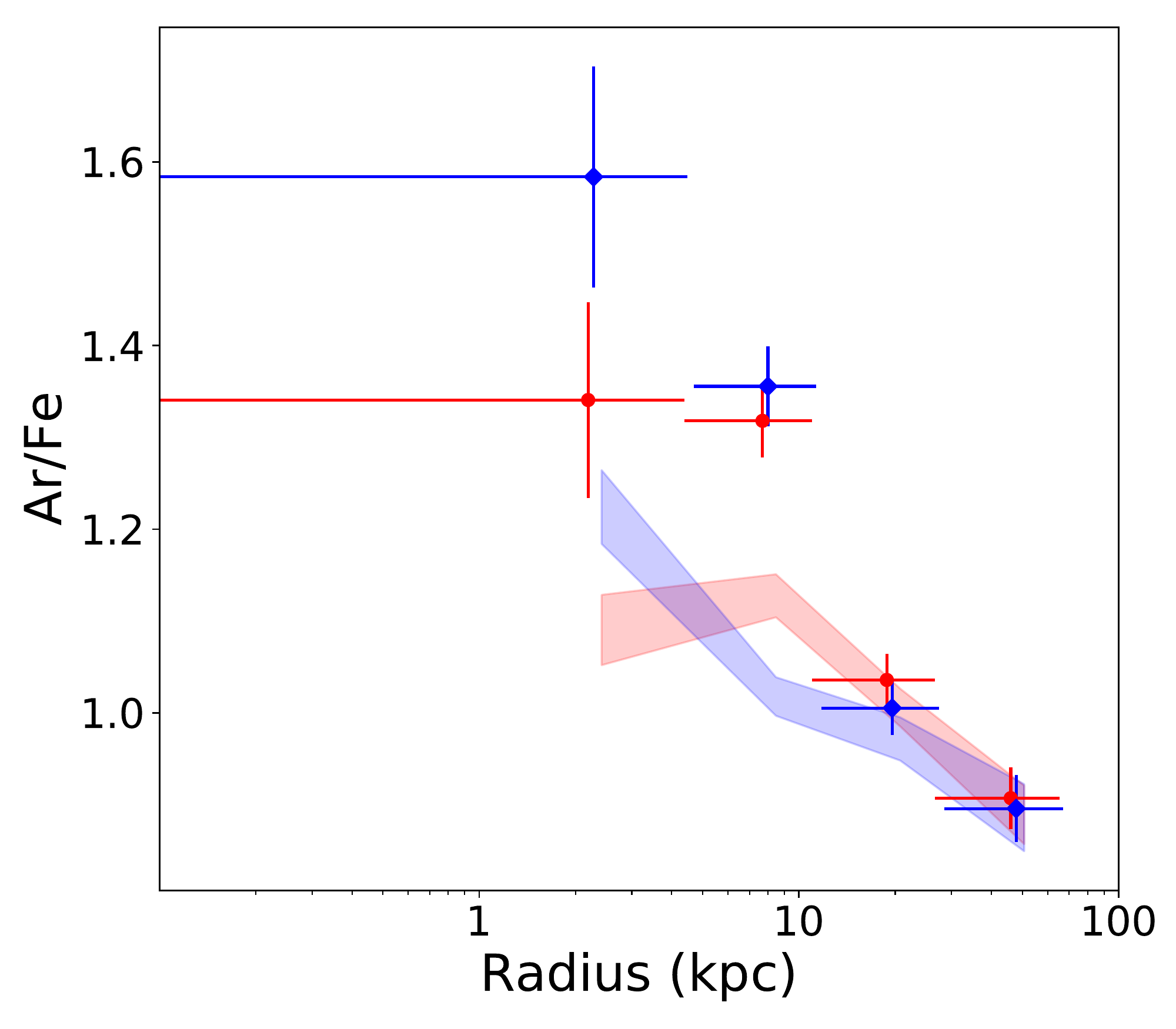}
\caption{A comparison of the projected (filled areas) and deprojected (data points) profiles of temperature, density, Fe, Si, S, Mg, Ca and Ar abundances and the Ar/Fe abundance ratio, obtained using the \textit{vgadem} (red) and 2T \textit{vapec} (blue) models. All errors are quoted at the 68.27\% (1$\sigma$) confidence level based on $\chi^2_{min}$. The blue data points have been shifted slightly on the X-axis for clarity.}
\label{fig:deproj_vgadem_vs_2Tvapec}
\end{figure*}

Projection of X-ray emission along the line of sight can blur out actual variations in the temperature and metal-abundance distribution, and
can sometimes also create artificial features. In order to get rid of these effects and investigate the drops as accurately as possible, it may be essential to deproject the spectra. As the first step, we deprojected the annular spectral regions using the \textit{dsdeproj} tool developed by \citet{Russell2008}. The method is based on the assumption of spherical symmetry, which is reasonable for the Centaurus cluster (see also Sect. \ref{sec:systematics_n_limitations}). The deprojected spectra were analyzed in the same manner as described above for the projected profiles (see Sect. \ref{sec:proj_prof}). The results for the radial distributions of temperature, total number density, the Fe, Si, S, Mg, Ca and Ar abundances, and the Ar/Fe abundance ratio obtained using the two models are shown in Fig.~\ref{fig:deproj_vgadem_vs_2Tvapec} (red and blue data points) and Table~\ref{tab:results}. In the 2T \textit{vapec} model, we find that only the innermost shell requires two components; the three outer shells can be reasonably fitted with a single-temperature component (\textit{vapec}). This result is not surprising, as the apparent multi-temperature nature of the ICM due to projection effects is expected to be strongly reduced after deprojection. 

Compared to their projected counterparts, the deprojected profiles modeled with the 2T \textit{vapec} and \textit{vgadem} models show very similar trends. We note, however, that the deprojected Fe, Si, S and Mg drops are somewhat more pronounced when using the 2T \textit{vapec} model. This may be explained by the fact that the shape of the unresolved Fe-L complex is rather sensitive to the assumed temperature structure. A change in the Fe abundance can, in turn, affect the abundance of the elements that are close to the Fe-L complex.
The agreement found between our projected and deprojected profiles in the outermost annuli/shells is not surprising, as the latter are weakly affected by the deprojection methods. On the contrary, the largest differences are to be expected in the inner regions, where projection effects are more important. In fact, the Fe drop (spanning over the three innermost bins) appears much more significant in the deprojected profiles, in particular when the 2T \textit{vapec} model is assumed. 
In addition, the Ar/Fe ratio also increases towards the center, with the same significance as in the projected profiles. 

Since we have demonstrated that the choice of the particular multi-temperature model does not dramatically affect the shape of the metal abundance drops (in particular for Ar, i.e., our main element of interest), in the following we choose to use the \textit{vgadem} model, both for its simplicity (one less free parameter than the 2T \textit{vapec} model) and realism (a continuous multi-temperature distribution seems more plausible than only two distinct, independent gas phases). Moreover, the weaker Ar/Fe gradient reported in Fig.~\ref{fig:deproj_vgadem_vs_2Tvapec} for the \textit{vgadem} model makes our results more conservative.

\subsubsection*{Deprojection methods}
\label{sec:deproj_methods}

\begin{figure*}
\centering
  \includegraphics[width=.32\linewidth]{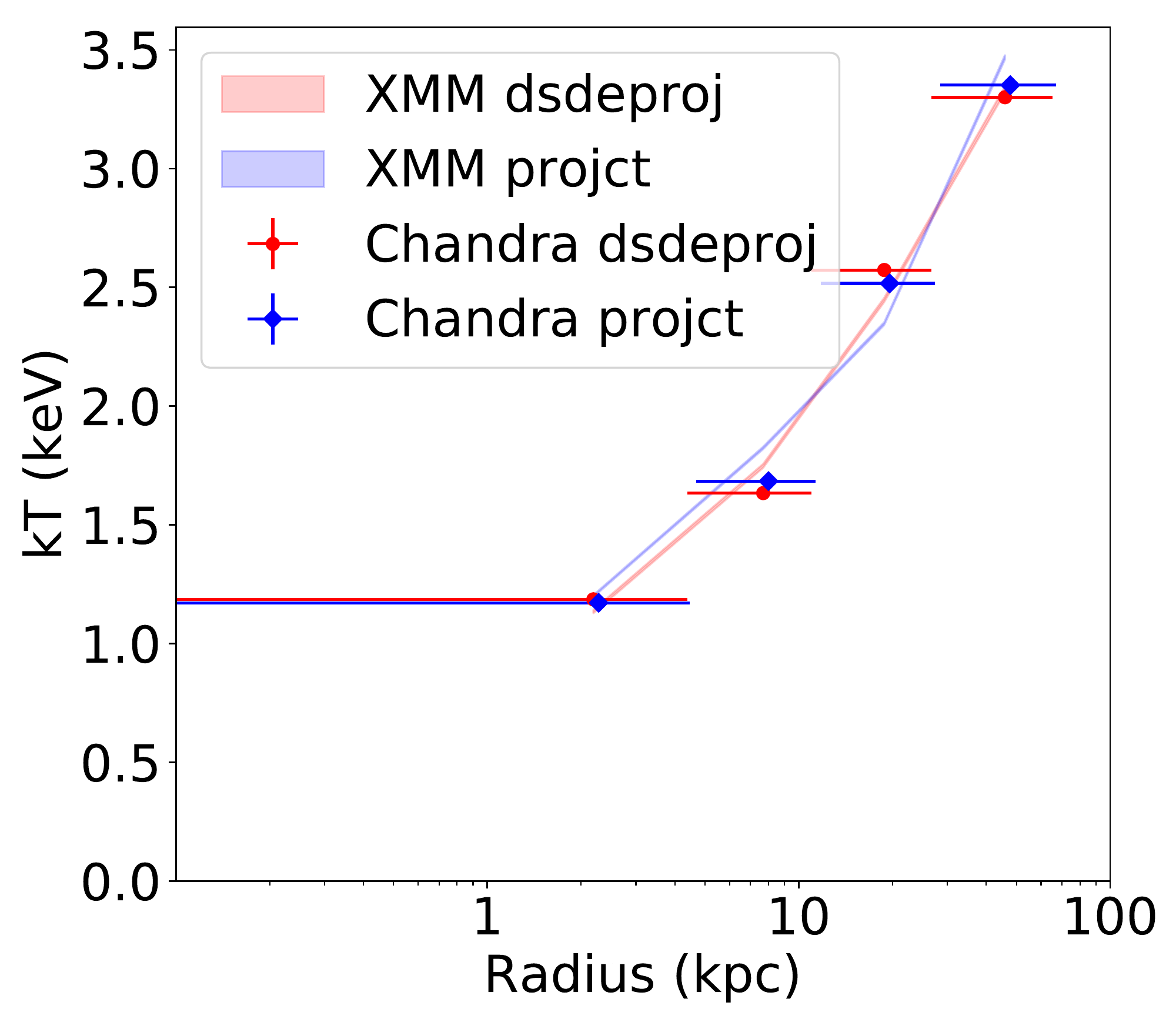}
  \includegraphics[width=.32\linewidth]{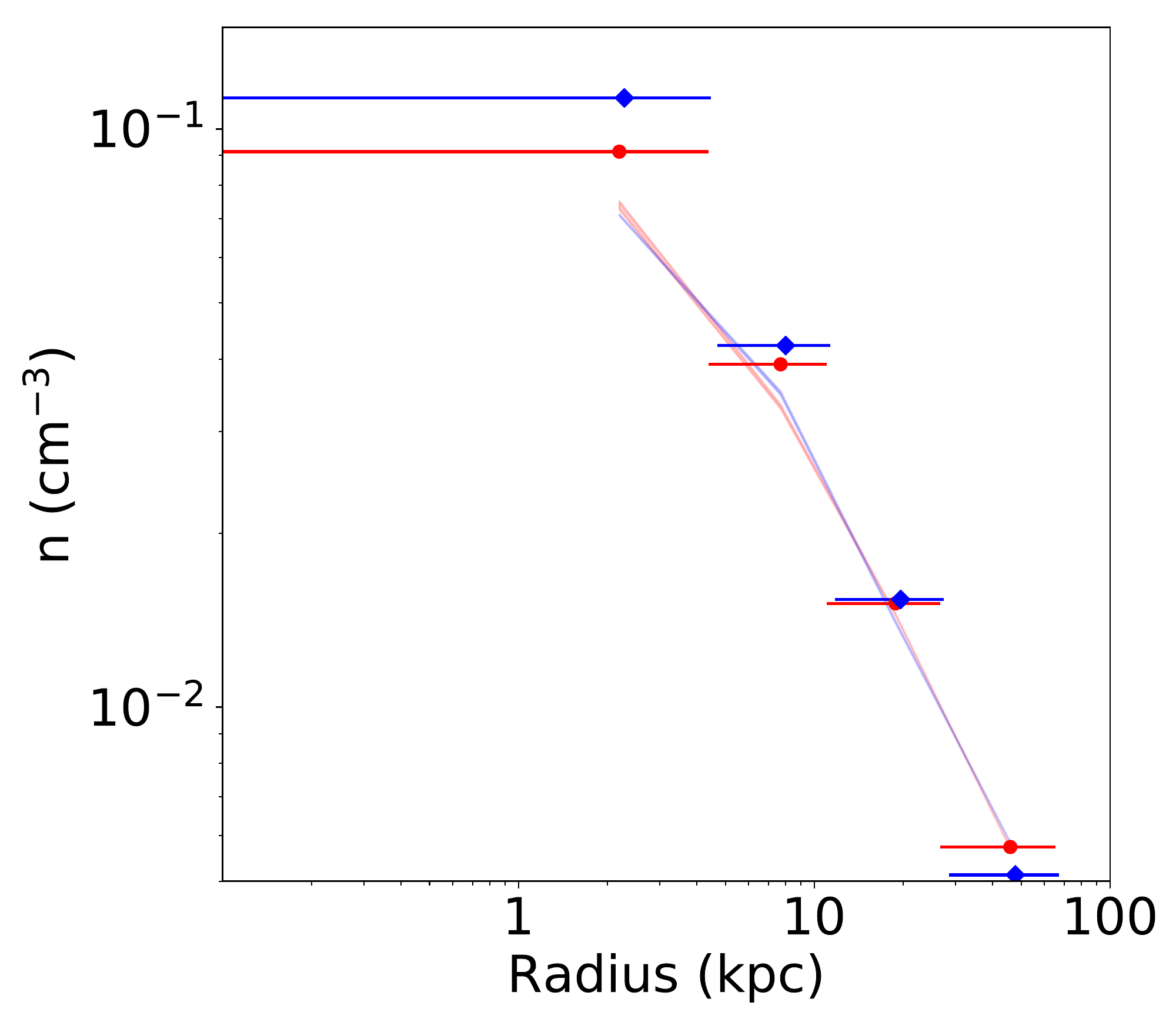}
  \includegraphics[width=.32\linewidth]{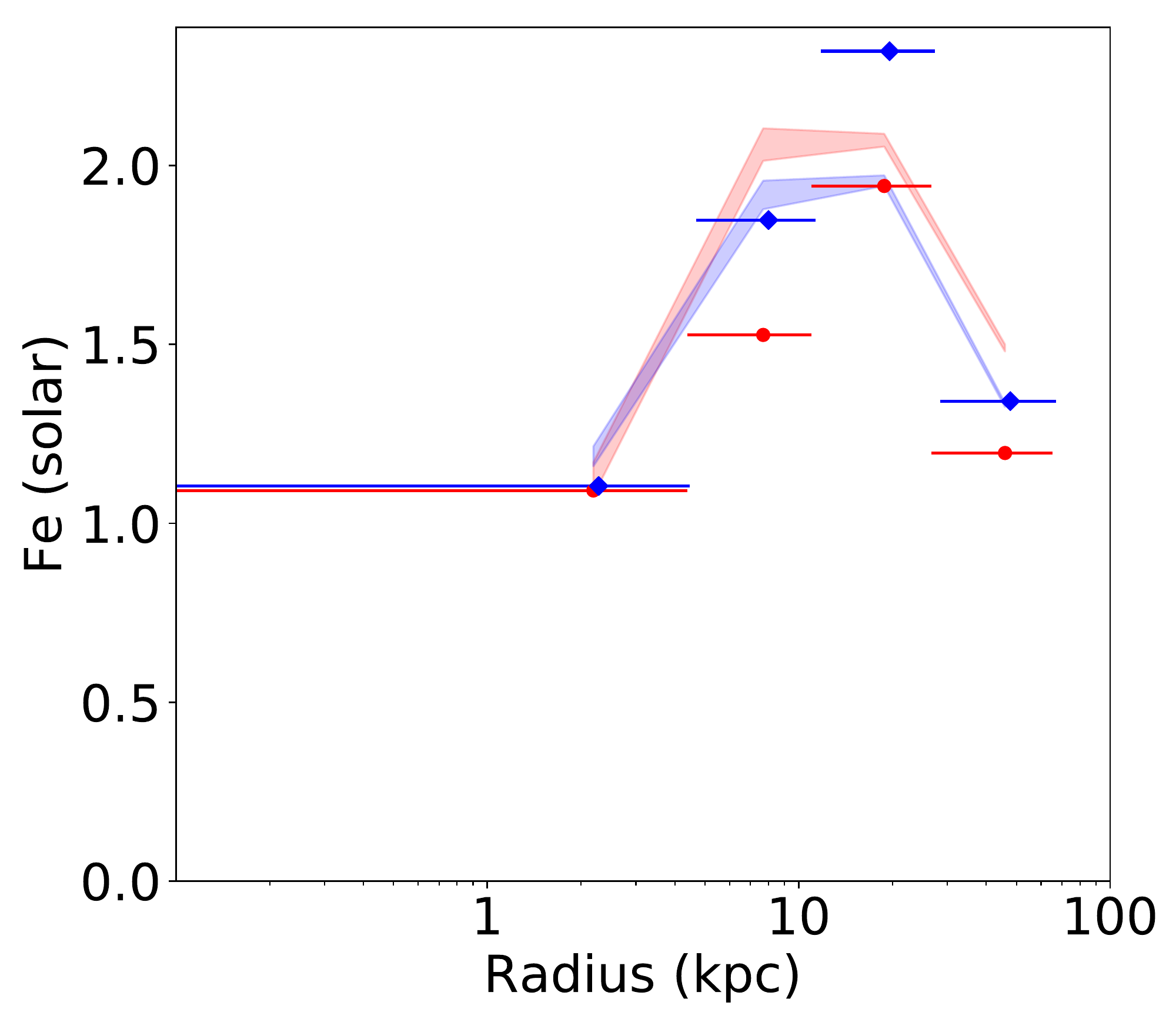}
  \includegraphics[width=.32\linewidth]{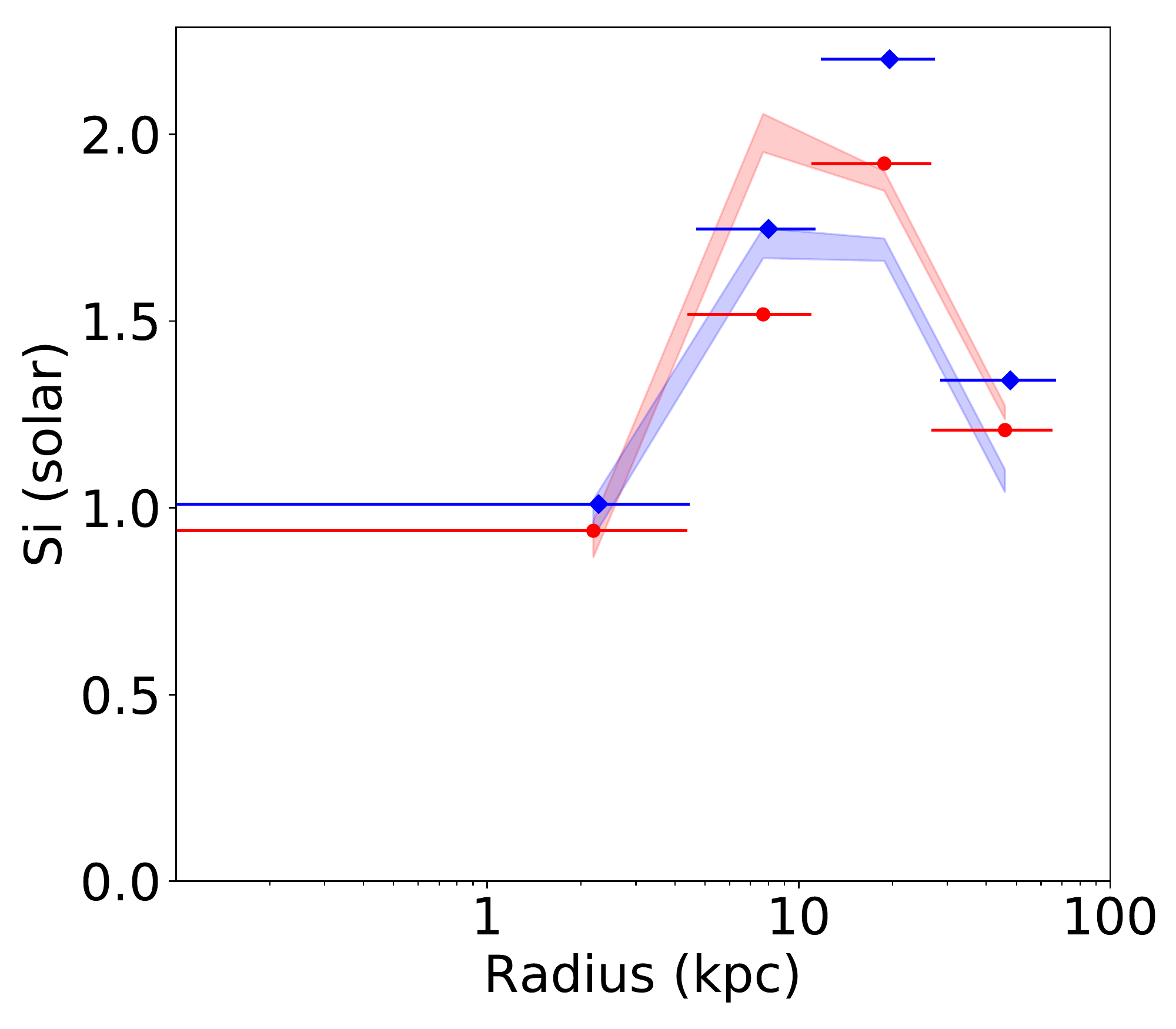}
  \includegraphics[width=.32\linewidth]{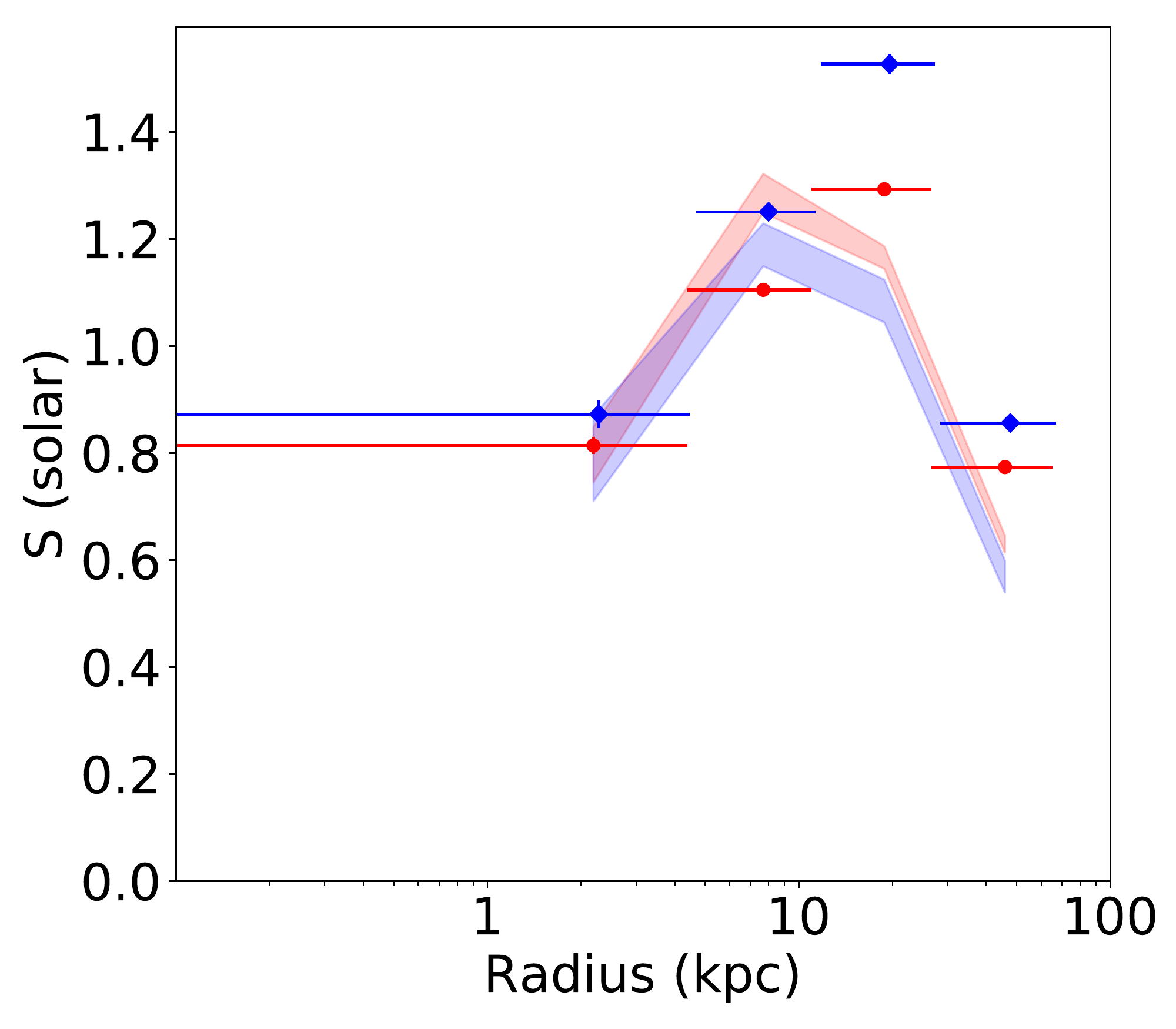}
  \includegraphics[width=.32\linewidth]{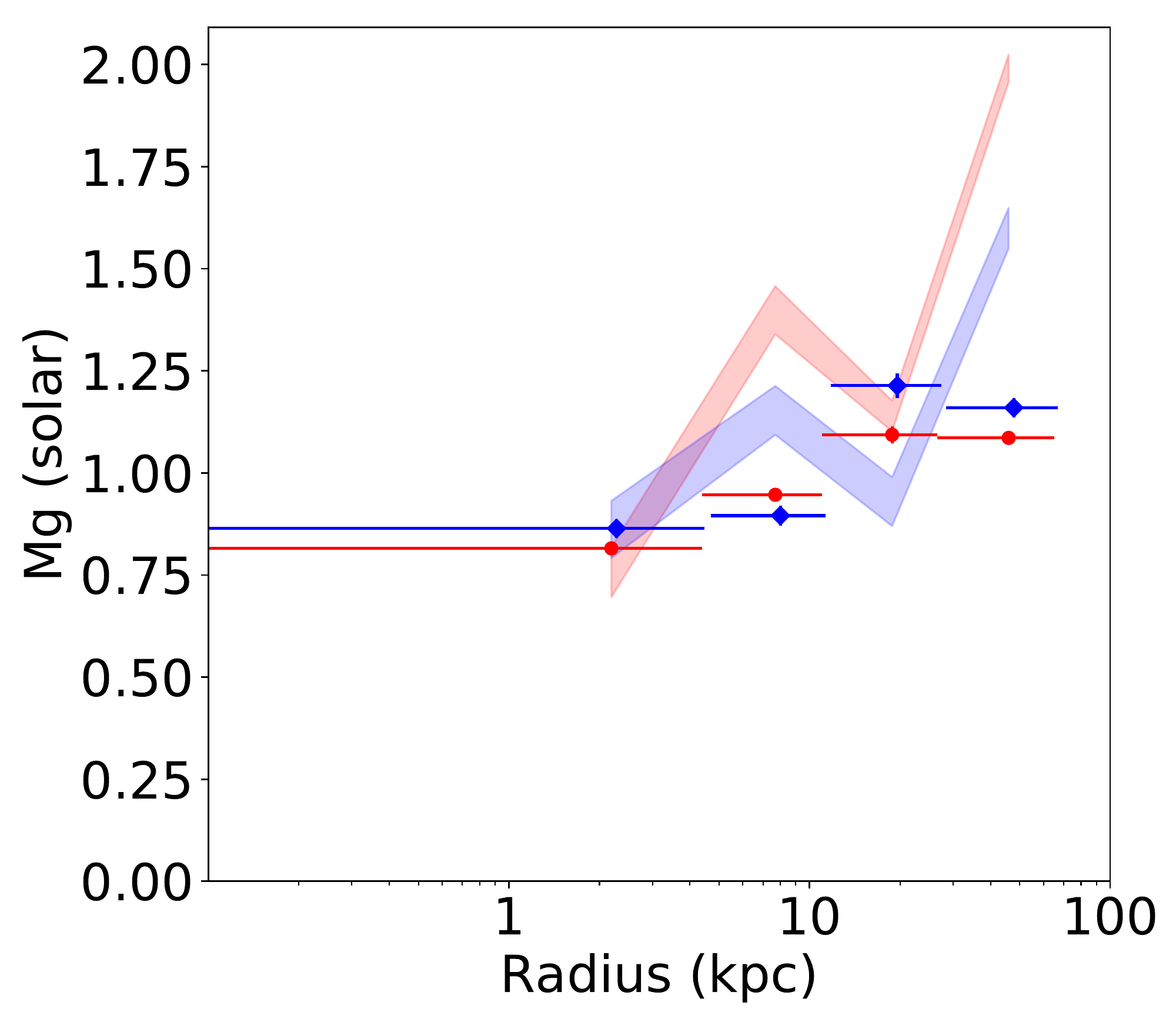}
  \includegraphics[width=.32\linewidth]{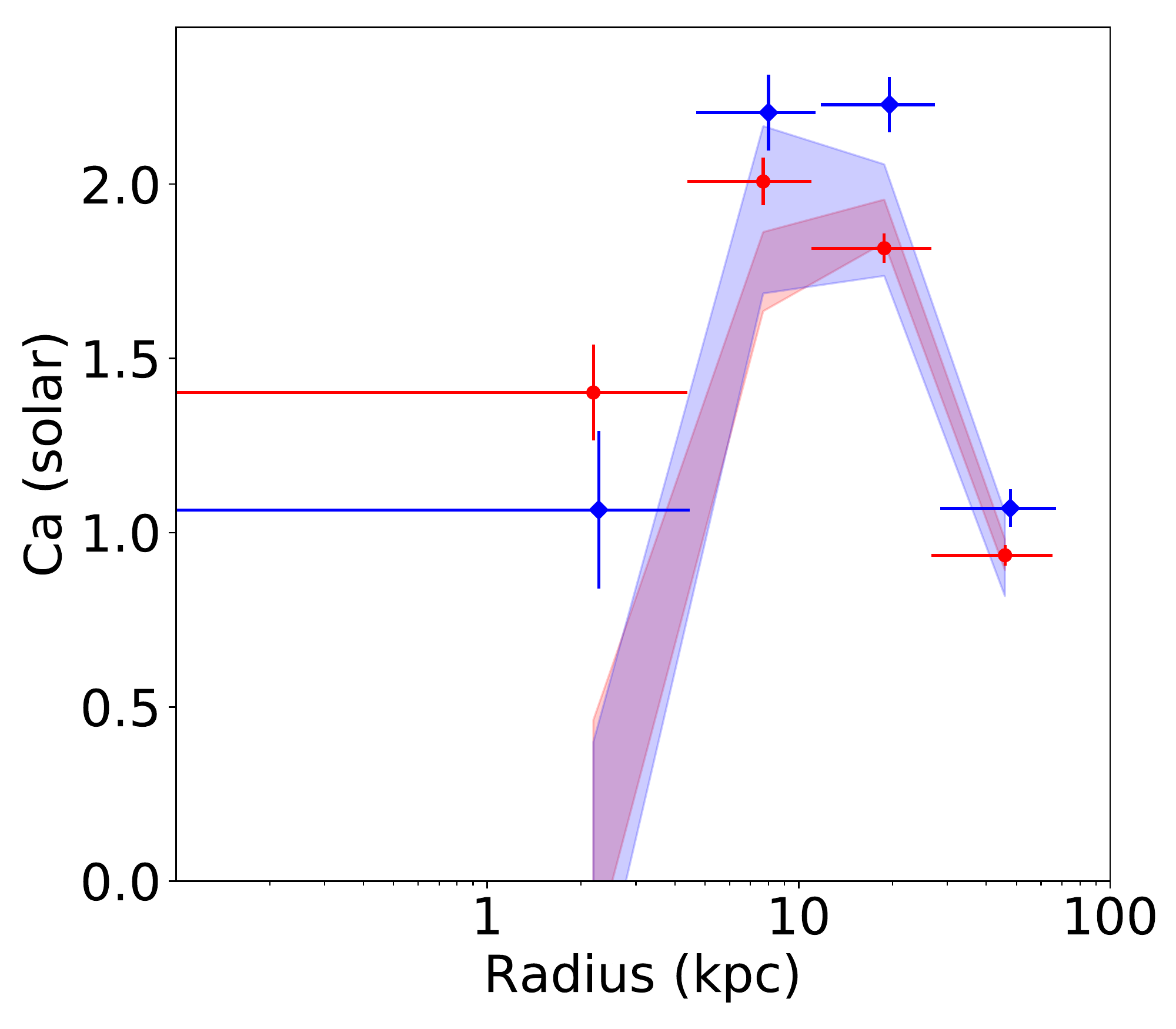}
  \includegraphics[width=.32\linewidth]{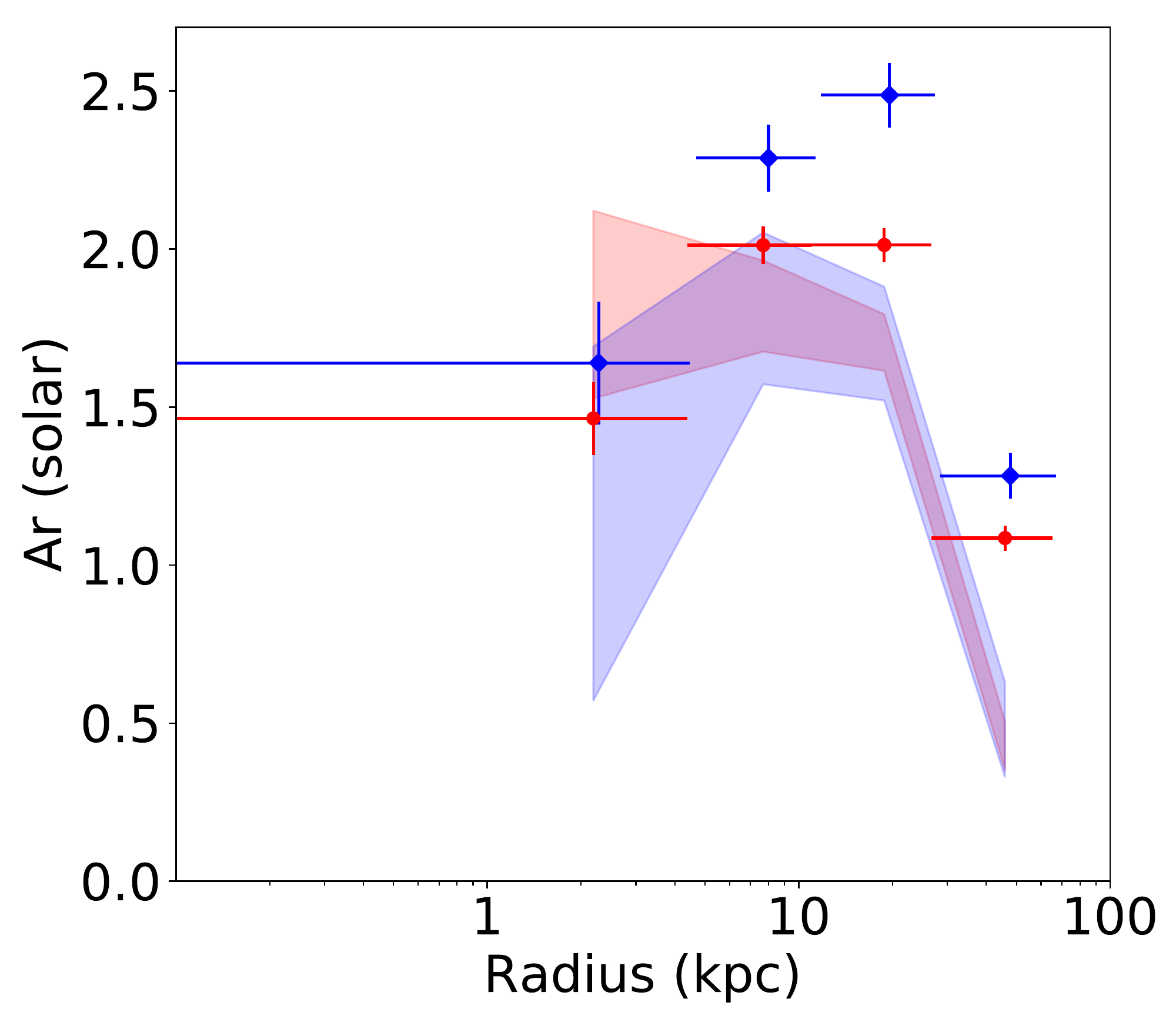}
  \includegraphics[width=.32\linewidth]{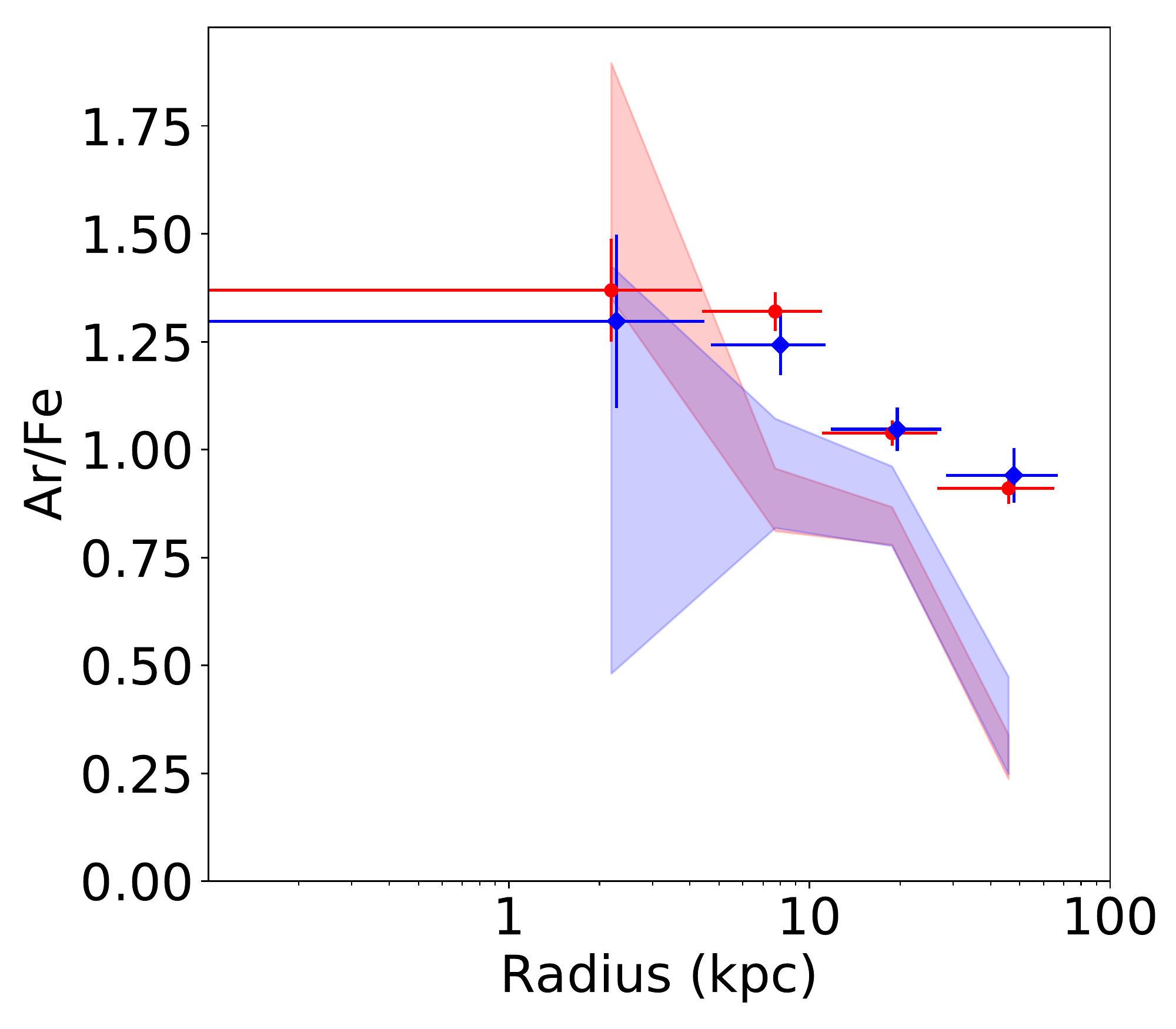}
\caption{A comparison of the \textit{Chandra} (data points) and \textit{XMM-Newton} (filled areas) temperature, density, Fe, Si, S, Mg, Ca and Ar abundances and the Ar/Fe abundance ratio profiles, obtained using the \textit{vgadem} model and deprojected using \textit{projct} (blue) and \textit{dsdeproj} (red). All errors are quoted at the 68.27\% (1$\sigma$) confidence level based on $\chi^2_{min}$. The blue data points have been shifted slightly on the X-axis for clarity.}
\label{fig:XMM_vs_Chandra_deproj_projct_vs_dsdeproj}
\end{figure*}

To test the sensitivity  of our results to different approaches to spectral deprojection, we compared our present approach (\textit{dsdeproj}) with measurements obtained using the \textit{projct} model in XSPEC. In the former method, for each shell the contributions of all the outer shells are subtracted and then each of the resulting deprojected spectra is fitted individually. The latter method involves forward fitting and the projected spectra of all the shells are fitted simultaneously to obtain the deprojected parameters. A comparison of the resulting profiles obtained using \textit{dsdeproj} and \textit{projct} is shown in Table~\ref{tab:results} and Fig.~\ref{fig:XMM_vs_Chandra_deproj_projct_vs_dsdeproj} (red and blue data points). 

The two deprojection methods yield rather similar metal abundance profile slopes. We note, however, that using \textit{projct} brings the value of the maximum abundance for Ca to better agreement with the $\sim$10 kpc extent of the Fe, Si, S and Mg drops. Furthermore, we find significantly higher abundances when using the \textit{projct} method. While this source of uncertainties has no impact for the purpose of this study, we caution that it may be more critical if one wants to constrain the integrated abundances or abundance ratios of some specific elements in the ICM \citep[e.g.,][]{Simionescu2018,Mernier2018b}. Since \textit{projct} involves a simultaneous fit of all the projected spectra, the derived uncertainties are higher for \textit{projct} than for \textit{dsdeproj}. Therefore, although the magnitude of the drops in the best-fit values of the metal abundances are larger for \textit{projct} than for \textit{dsdeproj}, the significance of the drops is still higher for \textit{dsdeproj} (41$\sigma$ vs. $30\sigma$ for Fe). The inter-dependency of the shell parameters in \textit{projct} is also known to produce potentially unstable measurements \citep{Russell2008}. However, the two deprojection techniques use different but equally valid approaches and the differences in the obtained results are likely to reflect the level of systematic uncertainties. For this reason, we show the results obtained with both methods. Because the implementation of \textit{dsdeproj} allows us to perform narrow band fits (see Sect. \ref{sec:narrow_band_analys}) more easily, we decided to use \textit{dsdproj} as our reference deprojection method.

\subsection{Comparison with \textit{XMM-Newton}}
\label{sec:chandra_xmm}

In addition to comparing the results between \textit{dsdeproj} and \textit{projct}, Table~\ref{tab:results} and Fig.~\ref{fig:XMM_vs_Chandra_deproj_projct_vs_dsdeproj} also compare the above \textit{Chandra}/ACIS-S (density, temperature, metal abundance and Ar/Fe ratio) profiles to the ones obtained with \textit{XMM-Newton}/EPIC (red and blue filled areas).

We find that, globally, the \textit{XMM-Newton} results are consistent (difference $<3\sigma$) with their corresponding \textit{Chandra} counterparts. The agreement is also in line with the analysis of \citet{Schellenberger2015}, who found that the temperatures measured by ACIS and EPIC tend to converge when the gas is cooler. We note, however, a slight but significant tension between the two data sets in some specific cases -- in particular in the second outermost shell when using \textit{projct}. We speculate that these discrepancies may be due to a combination of remaining calibration issues and artifacts propagated when \textit{projct} is used. Nevertheless, in both cases the deprojected EPIC spectra clearly reveal the presence of central drops for the Fe, Si, S, Mg and Ca abundances.  

Despite the formal consistency reported above, we note that the uncertainties of the \textit{XMM-Newton} abundances of Ar and Ca are much larger than the uncertainties of the \textit{Chandra} measurements. This point is discussed in more detail in Sect. \ref{sec:narrow_band_analys}.

\subsection{Broad-band versus narrow-band fits}
\label{sec:narrow_band_analys}

Whereas fitting the spectra using the full energy band  (i.e., typically $\sim$0.5-7 keV for {\it Chandra} ACIS) is the standard procedure for deriving the best-fit parameters, this approach may bias the metal abundance measurements. In fact, since the abundance of a given element is directly proportional to the equivalent width of its line (namely, the ratio between the total line flux and the continuum flux at the centroid energy of the line), incorrect determination of the continuum in the vicinity of the line would lead to incorrect measurement of the corresponding elemental abundance. Because no instrument is calibrated with infinite precision, some slight imperfections of the effective area may produce these local over- or underestimates of the continuum. In the case of very deep observations, that is, when statistical uncertainties become smaller than the above systematic effect, such a bias may become highly significant \citep[e.g.,][]{Mernier2015,Simionescu2018}. 

\begin{figure*}
\centering
  \includegraphics[width=.33\linewidth]{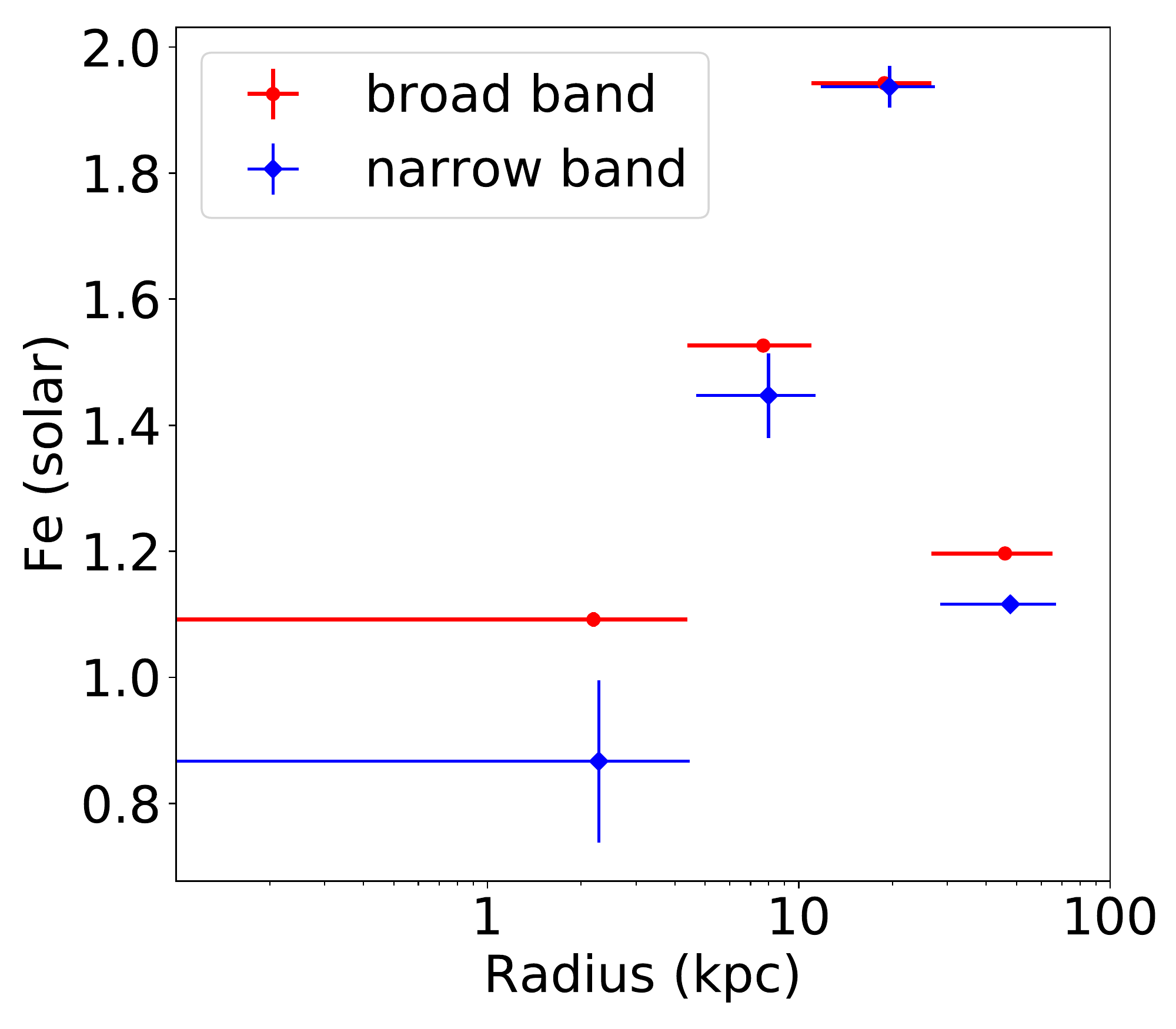}
  \includegraphics[width=.33\linewidth]{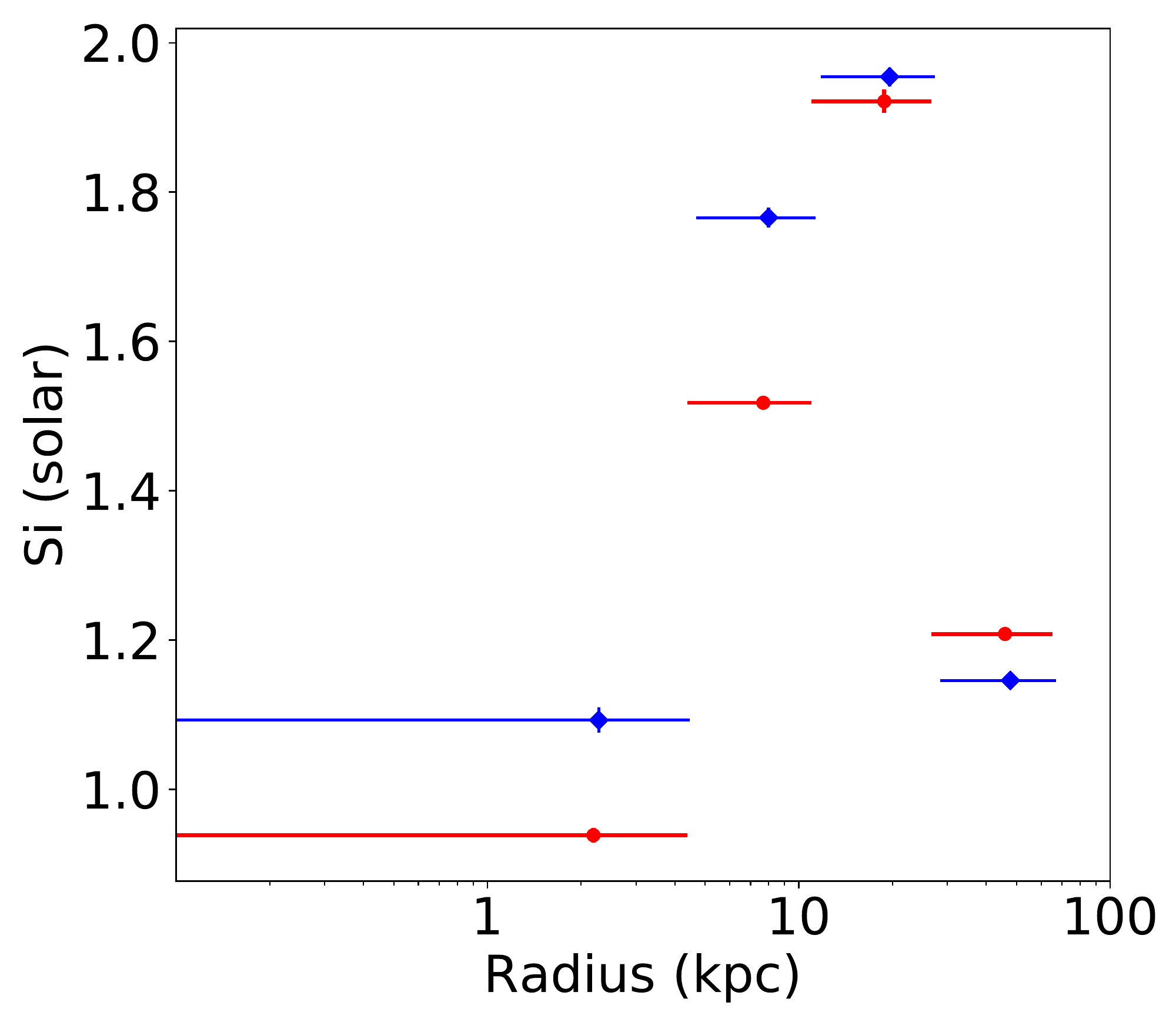}
  \includegraphics[width=.33\linewidth]{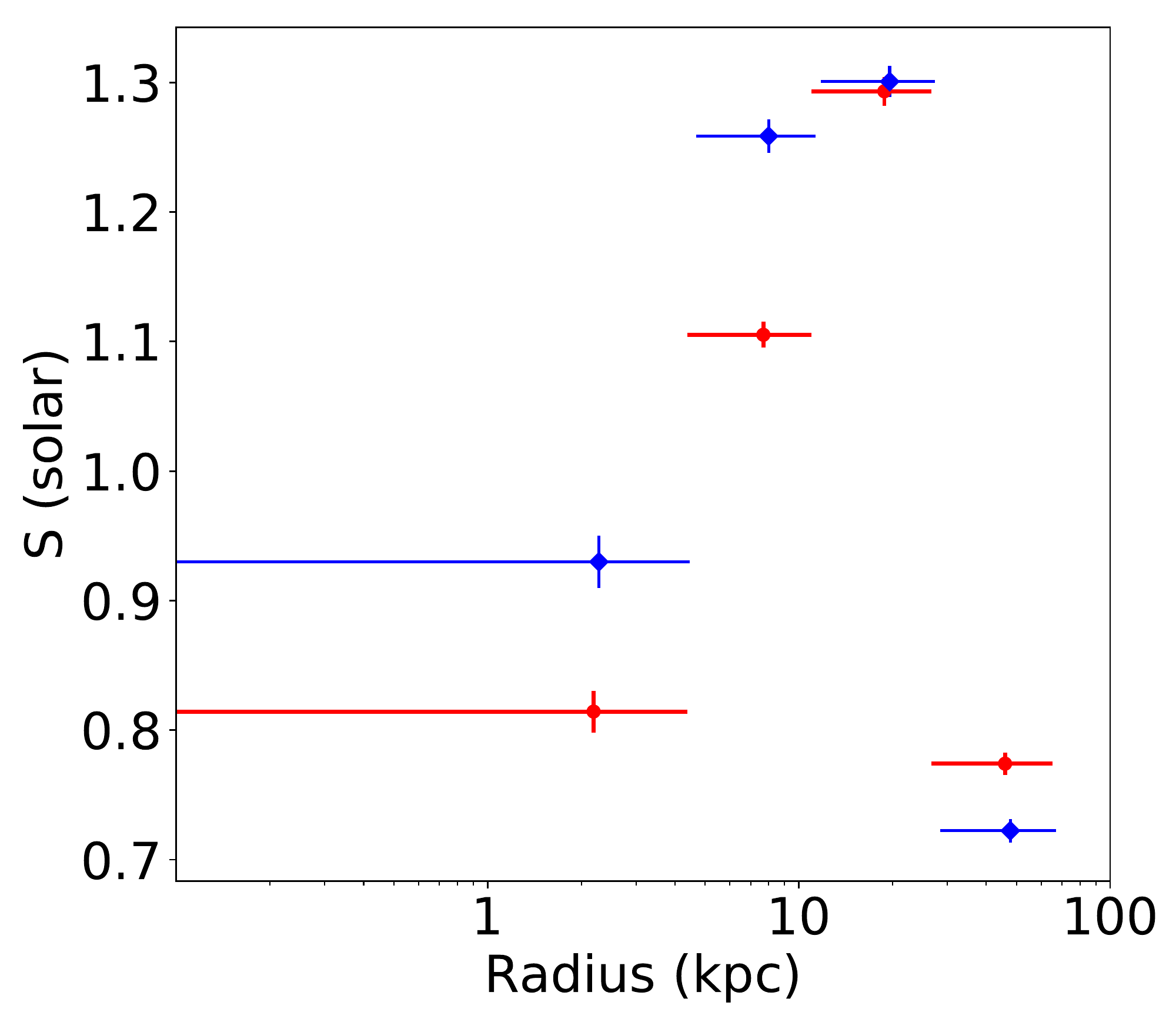}
  \includegraphics[width=.33\linewidth]{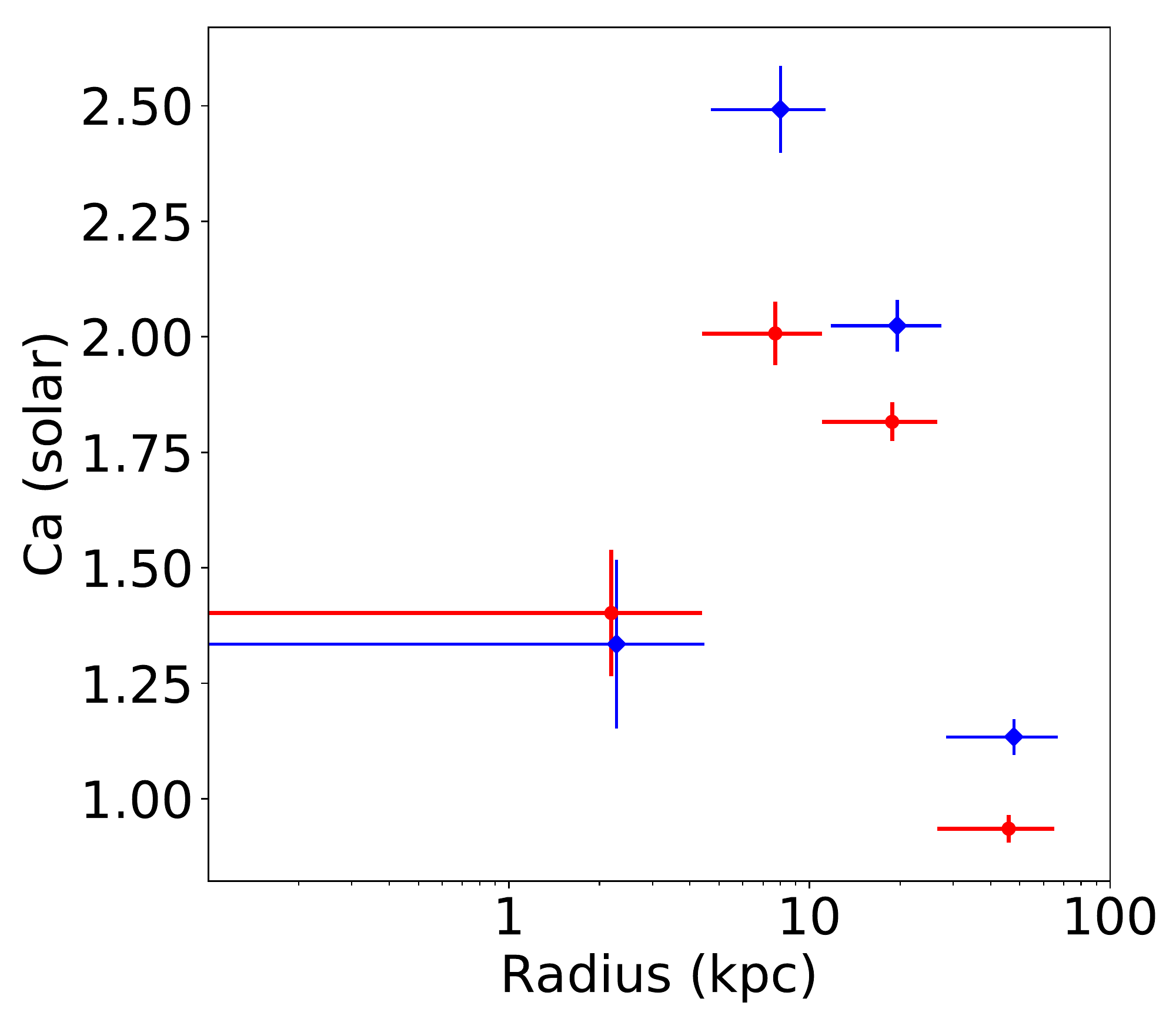}
  \includegraphics[width=.33\linewidth]{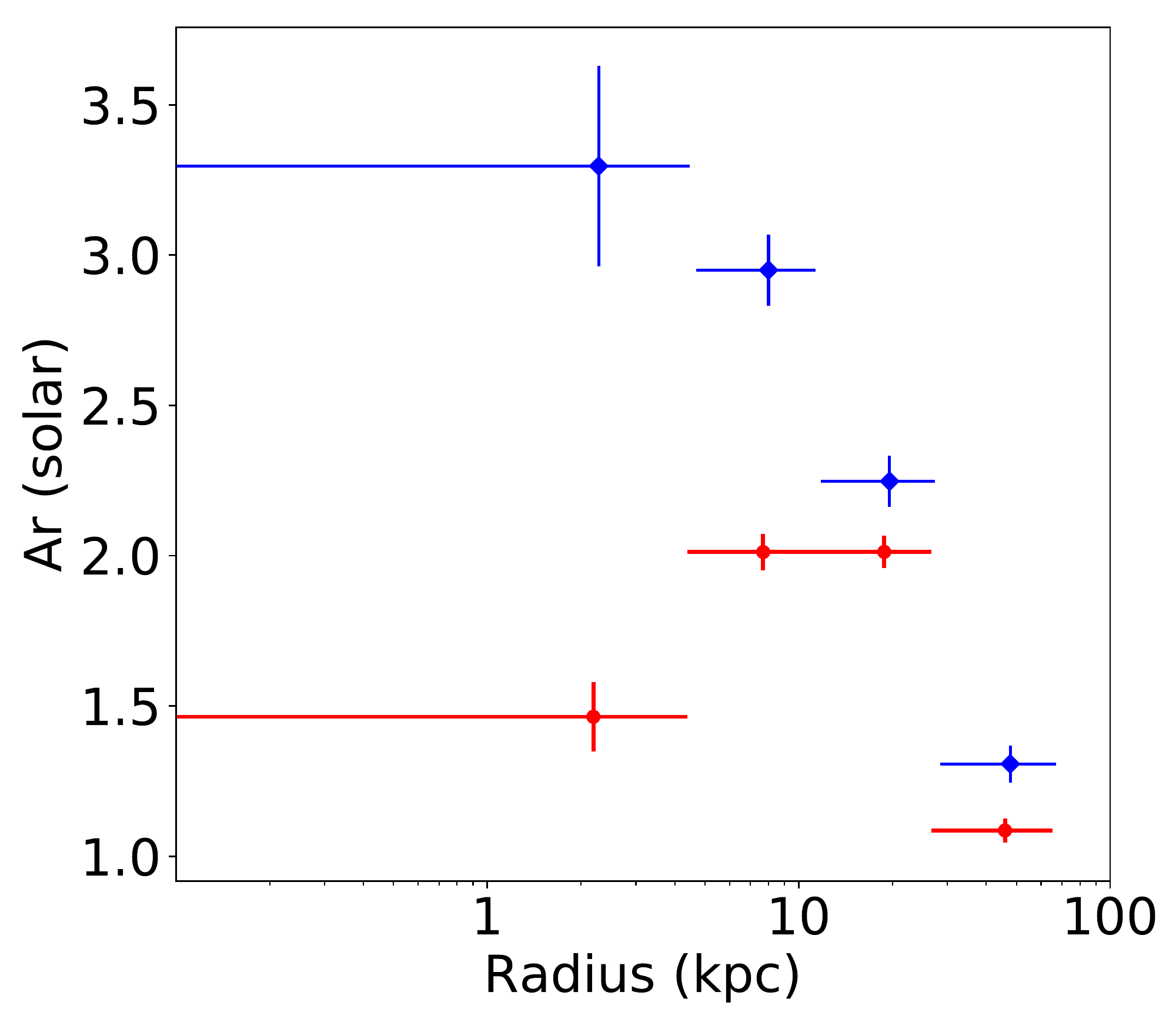}
  \includegraphics[width=.33\linewidth]{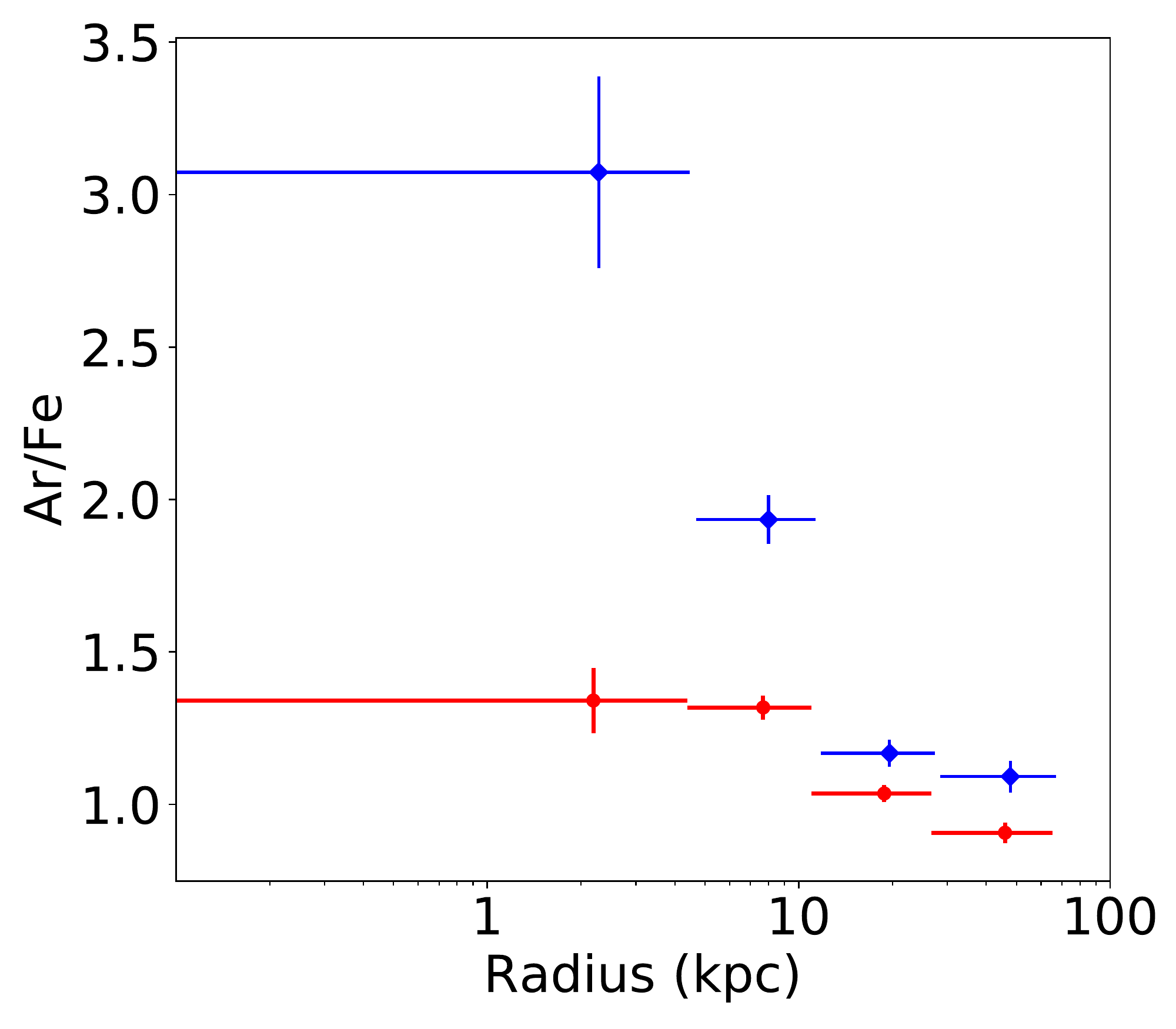}

\caption{Deprojected profiles of Fe, Si, S, Ca, and Ar abundances, and the Ar/Fe (Fe taken from the broad-band fits) ratio, using broad- (red) and narrow-band
(blue) fits, obtained using the \textit{vgadem} model and the \textit{dsdeproj} deprojection method. All errors are quoted at the 68.27\% (1$\sigma$) confidence level based on $\chi^2_{min}$. The blue data points have been shifted slightly on the X-axis for clarity.}
\label{fig:narrow_vs_broadband}
\end{figure*}

A good way to minimize such biases is to refit the spectra between several local energy bands, each of them including the K-shell emission line(s) used to derive the abundance of a given element as well as  part of the surrounding continuum. In these `narrow-band' fits (by contrast to the above `broad-band' approach), the only parameters that were kept free were the (local) normalization and the abundance of the considered element. All the other parameters were fixed to the best-fit value of the broad-band fit. A comparison of the resulting narrow-band fit profiles with the ones obtained using broad-band fits (assuming a \textit{vgadem} model and using \textit{dsdeproj}) is shown in Table~\ref{tab:results} and Fig.~\ref{fig:narrow_vs_broadband}. We used the 1.5-2.8 keV energy range for Si and S, 2.9-3.5 keV for Ar, 3.3-4.3 keV for Ca, and 5.0-7.0 keV for Fe.

The narrow-band Fe profile in Fig.~\ref{fig:narrow_vs_broadband} is consistent with the broad-band profile and also shows a drop in the innermost two bins (at $r\lesssim$10 kpc). The narrow-band fit results in larger statistical uncertainties on the Fe abundance, because it only includes the Fe-K lines and excludes the Fe-L complex. Despite the larger error bars on the measurements, the Fe drop in the narrow-band case is significant at the $>8\sigma$ level. 
Similarly to the broad-band profiles, the narrow-band profiles of Si, S, and Ca also show significant central drops (at the $\sim41$, 16, and 6$\sigma$ levels, respectively). Importantly, the narrow-band abundances of all the elements other than Fe are significantly higher than their broad-band values, especially in the innermost two bins. This indicates an overestimation of the continuum emission in the $\sim$1.5-5 keV range when the spectra are fitted over the entire 0.5-7 keV band.

Interestingly, the Ar abundance profile determined using narrow-band fits (Fig.~\ref{fig:narrow_vs_broadband}, bottom left), differs even more from the corresponding broad-band results than the other profiles reported above. As reported in Sect. \ref{sec:deproj_prof}, the broad-band Ar profile shows a $2\sigma$ hint of a drop in the innermost bin, while the highest Ar value peaks between 4 and 27 kpc. However, the continuum around the Ar line in the innermost bin was found to be slightly overestimated. We also confirm the overestimation of the continuum using \textit{projct}. After applying the narrow-band fit correction, the centrally decreasing behavior of Ar completely disappears and becomes either flat or increasing towards the center. Perhaps even more importantly, the central increase of the Ar/Fe ratio \footnote{In the Ar/Fe profile shown in Fig.~\ref{fig:narrow_vs_broadband}, we choose to adopt the Fe values from their broad-band measurements because the Fe-L complex provides additional constraints while not deviating significantly from the narrow-band values. We note that, when adopting the narrow-band Fe values in the Ar/Fe profile, the central increase remains firmly detected.} reported earlier is now strongly enhanced ($>3\sigma$ comparing the innermost two bins; $>6\sigma$ when comparing the innermost and outermost bins; Fig.~\ref{fig:narrow_vs_broadband}, bottom right).

\begin{figure*}
\centering
  \includegraphics[width=.7\linewidth]{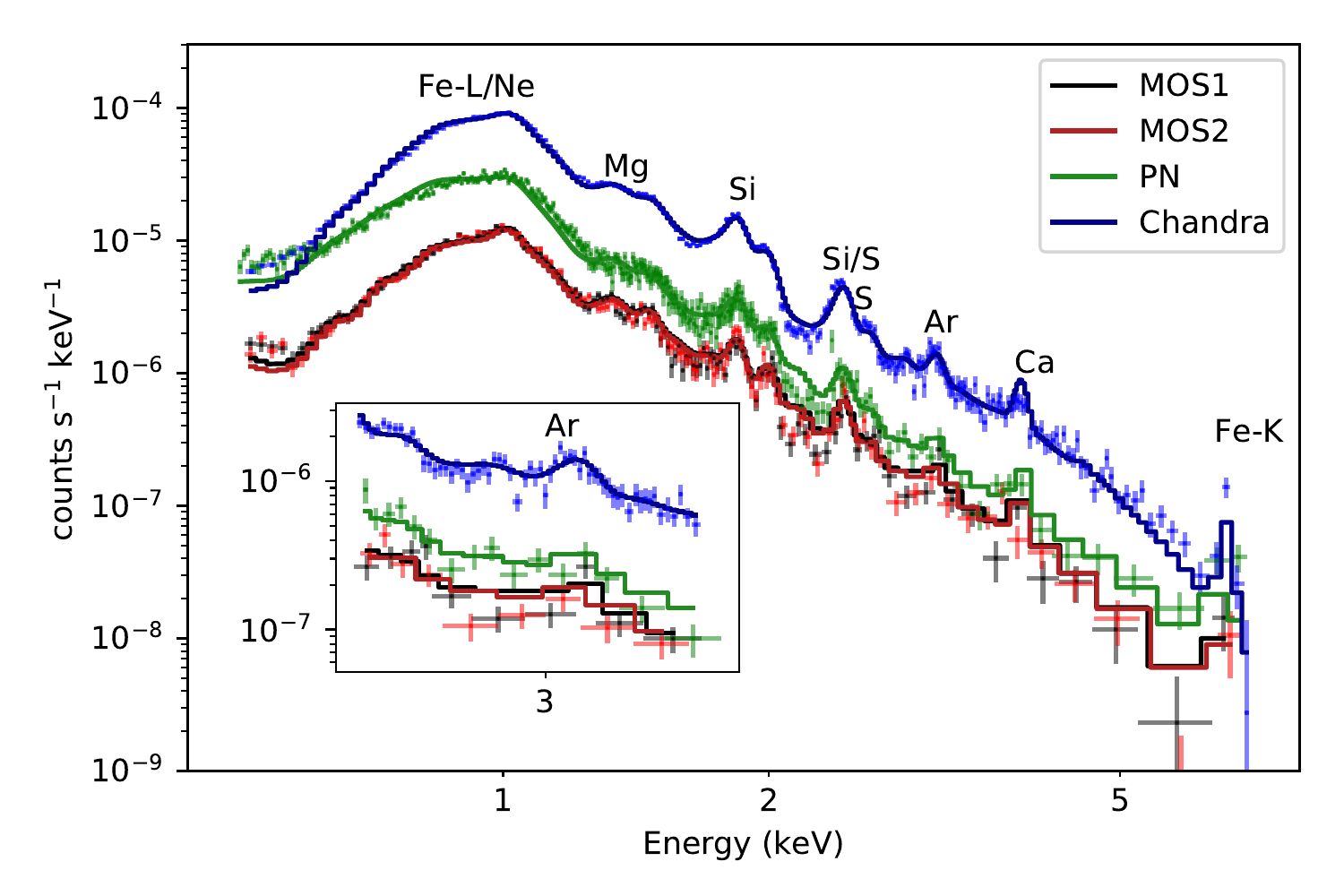}
\caption{A comparison of the deprojected \textit{Chandra} (blue) and the \textit{XMM-Newton} MOS1 (black), MOS2 (red), and PN (green) spectra of the innermost shell; the most important emission lines have been marked. The {\it Chandra} spectrum has been scaled up by a factor of five for viewing purposes. The inset shows the zoomed-in spectra around the Ar line. It can be seen that a slight over/underestimation of the continuum in the broad-band fit can lead to an under/overestimation of the abundances, especially for the weak lines. It is also clearly seen that the \textit{XMM-Newton} spectra around the high-energy lines 
(Ar, Ca and Fe-K) are very noisy and are not as resolved as in the \textit{Chandra} spectrum and therefore the \textit{XMM-Newton} spectra are not found suitable for the narrow-band analyses.}
\label{fig:anu1_deproj_Chandra_XMM}
\end{figure*}

Given the rather deep ($\sim$150 ks) \textit{XMM-Newton} observations that are available, and motivated by our previous comparison in Sect. \ref{sec:chandra_xmm}, we also attempted to perform narrow-band fits in the deprojected \textit{XMM-Newton} spectra. Regrettably, very few additional constraints could be obtained for Ar, Ca, and Fe using that approach (see also Sect. \ref{sec:chandra_xmm}). In fact, the limited number of counts for each of the EPIC instruments, coupled with the spectral grouping requirements of \textit{dsdeproj}, make narrow-band fits of their deprojected spectra around the Ar line very difficult. This is illustrated further in Fig.~\ref{fig:anu1_deproj_Chandra_XMM}, where beyond $\sim$2 keV the combined deprojected spectra of ACIS-S is better resolved than the combined deprojected spectra of EPIC MOS\,1, MOS\,2, and pn. In theory, better constraints from the \textit{XMM-Newton} data would be possible if the three EPIC spectra were stacked instead of being fitted simultaneously; however this method is not advised as it may introduce unexpected uncertainties from the combined response and effective area.

For this reason, we choose to rely primarily on the deeper \textit{Chandra} observations, as they allow us to keep all the systematic effects mentioned above under control.

\begin{table*}
\caption{Best-fit values for the temperature, density, Fe, Si, S, Mg, Ca, and Ar abundances, and the Ar/Fe ratio (for the narrow band analysis, the Fe abundance is taken from the broad-band fit), and their associated uncertainties (quoted at the 68.27\% or 1$\sigma$ level) obtained from the various spectral fitting and deprojection methods employed in the paper. We note that here we quote the mean temperature for the \textit{vgadem} model and the higher temperature for the 2T \textit{vapec} model.}
\label{tab:results}
\centering
\footnotesize
\begin{tabular}{cccccccccc}
\hline
Radial range & kT & n &  Fe & Si  & S & Mg & Ca & Ar & Ar/Fe \\
(kpc). & (keV) & (10$^{-2}$ cm$^{-3}$) &  (solar) & (solar)  & (solar) & (solar) & (solar) & (solar) &   \\
\hline
\hline\multicolumn{10}{c}{\textbf{\textit{Chandra} Projected }}     \\
\hline\multicolumn{10}{c}{Model: \textit{vgadem}}     \\
\hline
0--4.4 & 1.600$\pm$0.002 & 10.78$\pm$0.04     & 1.53$\pm$0.01 & 1.37$\pm$0.01 & 1.05$\pm$0.01 & 1.05$\pm$0.01 & 1.53$\pm$0.06 & 1.67$\pm$0.06 & 1.09$\pm$0.04 \\
4.4--11.0 & 1.945$\pm$0.002 &  4.02$\pm$0.01     & 1.86$\pm$0.01 & 1.79$\pm$0.01 & 1.24$\pm$0.01 & 1.07$\pm$0.01 & 1.91$\pm$0.04 & 2.10$\pm$0.04 & 1.13$\pm$0.02 \\
11.0--26.6 & 2.648$\pm$0.003 & 1.463$\pm$0.001    & 1.90$\pm$0.01 & 1.86$\pm$0.01 & 1.24$\pm$0.01 & 1.18$\pm$0.01 & 1.64$\pm$0.03 & 1.92$\pm$0.04 & 1.01$\pm$0.02 \\
26.6--65.3 & 3.278$\pm$0.007 & 0.4885 $\pm$0.0004 & 1.21$\pm$0.01 & 1.23$\pm$0.01 & 0.78$\pm$0.01 & 1.10$\pm$0.01 & 0.92$\pm$0.03 & 1.08$\pm$0.04 & 0.89$\pm$0.03 \\
\hline\multicolumn{10}{c}{Model: 2T \textit{vapec}}     \\
\hline
0--4.4 & 1.604$\pm$0.003 & 13.49$\pm$0.01  & 1.406$\pm$0.005 & 1.27$\pm$0.01 & 0.98$\pm$0.01 & 0.97$\pm$0.01 & 1.62$\pm$0.06 & 1.72$\pm$0.06 & 1.22$\pm$0.04 \\
4.4--11.0 & 1.992$\pm$0.003 & 5.428$\pm$0.004 & 1.977$\pm$0.005 & 1.77$\pm$0.01 & 1.25$\pm$0.01 & 1.07$\pm$0.01 & 1.92$\pm$0.04 & 2.01$\pm$0.04 & 1.02$\pm$0.02 \\
11.0--26.6 & 2.568$\pm$0.004 & 2.042$\pm$0.005 & 1.90$\pm$0.01   & 1.81$\pm$0.01 & 1.22$\pm$0.01 & 1.16$\pm$0.02 & 1.62$\pm$0.03 & 1.84$\pm$0.04 & 0.97$\pm$0.02 \\
26.6--65.3 & 3.204$\pm$0.007 & 0.683$\pm$0.002 & 1.23$\pm$0.01   & 1.22$\pm$0.01 & 0.78$\pm$0.01 & 1.13$\pm$0.02 & 0.93$\pm$0.03 & 1.09$\pm$0.04 & 0.89$\pm$0.04 \\
\hline
\hline\multicolumn{10}{c}{\textbf{\textit{Chandra} Deprojected }} \\
\hline\multicolumn{10}{c}{Deprojection Method: \textit{dsdeproj}}     \\
\multicolumn{10}{c}{Model: \textit{vgadem}}     \\
\hline
0--4.4 & 1.187$\pm$0.001 & 9.14$\pm$0.04   & 1.09$\pm$0.01   & 0.94$\pm$0.01 & 0.81$\pm$0.02 & 0.82$\pm$0.02 & 1.40$\pm$0.14 & 1.46$\pm$0.12 & 1.34$\pm$0.11 \\
4.4--11.0 & 1.634$\pm$0.001 & 3.918$\pm$0.003 & 1.53$\pm$0.01   & 1.52$\pm$0.01 & 1.11$\pm$0.01 & 0.94$\pm$0.02 & 2.01$\pm$0.07 & 2.01$\pm$0.06 & 1.32$\pm$0.04 \\
11.0--26.6 & 2.573$\pm$0.006 & 1.512$\pm$0.003 & 1.94$\pm$0.01   & 1.92$\pm$0.02 & 1.29$\pm$0.01 & 1.09$\pm$0.02 & 1.82$\pm$0.04 & 2.01$\pm$0.05 & 1.04$\pm$0.03 \\
26.6--65.3 & 3.301$\pm$0.008 & 0.573$\pm$0.001 & 1.20$\pm$0.01   & 1.21$\pm$0.01 & 0.77$\pm$0.01 & 1.09$\pm$0.02 & 0.93$\pm$0.03 & 1.09$\pm$0.04 & 0.91$\pm$0.03 \\
\hline\multicolumn{10}{c}{Deprojection Method: \textit{dsdeproj}}     \\
\multicolumn{10}{c}{Model: 2T \textit{vapec}}     \\
\hline
0--4.4 & 1.318$\pm$0.007 & 10.82$\pm$0.11  & 0.93$\pm$0.02   & 0.81$\pm$0.02 & 0.73$\pm$0.02 & 0.68$\pm$0.03 & 1.45$\pm$0.14 & 1.47$\pm$0.11 & 1.58$\pm$0.12 \\
4.4--11.0 & 1.600$\pm$0.001 & 5.78$\pm$0.01   & 1.40$\pm$0.01   & 1.36$\pm$0.01 & 0.99$\pm$0.01 & 0.67$\pm$0.02 & 1.91$\pm$0.07 & 1.90$\pm$0.06 & 1.36$\pm$0.04 \\
11.0--26.6 & 2.483$\pm$0.004 & 2.096$\pm$0.003 & 1.97$\pm$0.01   & 1.92$\pm$0.01 & 1.29$\pm$0.01 & 1.09$\pm$0.02 & 1.78$\pm$0.04 & 1.98$\pm$0.06 & 1.01$\pm$0.03 \\
26.6--65.3 & 3.196$\pm$0.005 & 0.794$\pm$0.001 & 1.23$\pm$0.01   & 1.21$\pm$0.01 & 0.77$\pm$0.01 & 1.13$\pm$0.02 & 0.94$\pm$0.03 & 1.10$\pm$0.04 & 0.89$\pm$0.04 \\
\hline
\multicolumn{10}{c}{Deprojection Method: \textit{projct}}     \\
\multicolumn{10}{c}{Model: \textit{vgadem}}     \\
\hline
0--4.4 & 1.173$\pm$0.002 & 11.32$\pm$0.12  & 1.104$\pm$0.003 & 1.01$\pm$0.02 & 0.87$\pm$0.03 & 0.86$\pm$0.02 & 1.07$\pm$0.23 & 1.64$\pm$0.19 & 1.30$\pm$0.20 \\
4.4--11.0 & 1.685$\pm$0.001 & 4.22$\pm$0.03   & 1.847$\pm$0.004 & 1.74$\pm$0.01 & 1.25$\pm$0.02 & 0.90$\pm$0.02 & 2.21$\pm$0.11 & 2.29$\pm$0.11 & 1.24$\pm$0.07 \\
11.0--26.6 & 2.517$\pm$0.003 & 1.536$\pm$0.005 & 2.319$\pm$0.005 & 2.20$\pm$0.02 & 1.53$\pm$0.02 & 1.21$\pm$0.03 & 2.23$\pm$0.08 & 2.49$\pm$0.10 & 1.05$\pm$0.05 \\
26.6--65.3 & 3.352$\pm$0.005 & 0.513$\pm$0.001 & 1.341$\pm$0.004 & 1.34$\pm$0.01 & 0.86$\pm$0.01 & 1.16$\pm$0.02 & 1.07$\pm$0.05 & 1.28$\pm$0.07 & 0.94$\pm$0.06 \\
\hline
\hline\multicolumn{10}{c}{\textbf{\textit{XMM-Newton} Deprojected }}  \\
\hline\multicolumn{10}{c}{Deprojection Method: \textit{dsdeproj}}     \\
\multicolumn{10}{c}{Model: \textit{vgadem}}     \\
\hline
0--4.4 & 1.14$\pm$0.01 & 7.38$\pm$0.12   & 1.12$\pm$0.05 & 0.92$\pm$0.05 & 0.80$\pm$0.05 & 0.76$\pm$0.06 & 0.13$\pm$0.33 & 1.82$\pm$0.30 & 1.62$\pm$0.27 \\
4.4--11.0 & 1.75$\pm$0.01 & 3.31$\pm$0.02   & 2.06$\pm$0.05 & 2.00$\pm$0.05 & 1.29$\pm$0.04 & 1.40$\pm$0.06 & 1.75$\pm$0.11 & 1.82$\pm$0.14 & 0.88$\pm$0.07 \\
11.0--26.6 & 2.45$\pm$0.01 & 1.445$\pm$0.004 & 2.07$\pm$0.02 & 1.88$\pm$0.03 & 1.17$\pm$0.02 & 1.14$\pm$0.04 & 1.89$\pm$0.06 & 1.70$\pm$0.09 & 0.82$\pm$0.04 \\
26.6--65.3 & 3.33$\pm$0.01 & 0.570$\pm$0.001 & 1.49$\pm$0.01 & 1.25$\pm$0.02 & 0.63$\pm$0.02 & 1.99$\pm$0.03 & 0.94$\pm$0.04 & 0.43$\pm$0.08 & 0.29$\pm$0.05 \\
\hline\multicolumn{10}{c}{Deprojection Method: \textit{projct}}     \\
\multicolumn{10}{c}{Model: \textit{vgadem}}     \\
\hline
0--4.4 & 1.62$\pm$0.27 & 7.10$\pm$0.02   & 1.19$\pm$0.03 & 0.97$\pm$0.05 & 0.79$\pm$0.08 & 0.86$\pm$0.07 & 0.00$\pm$0.40 & 1.13$\pm$0.56 & 0.95$\pm$0.47 \\
4.4--11.0 & 1.82$\pm$0.01 & 3.49$\pm$0.02   & 1.92$\pm$0.04 & 1.71$\pm$0.04 & 1.19$\pm$0.04 & 1.15$\pm$0.06 & 1.93$\pm$0.24 & 1.81$\pm$0.24 & 0.94$\pm$0.13 \\
11.0--26.6 & 2.35$\pm$0.01 & 1.40$\pm$0.01   & 1.96$\pm$0.02 & 1.69$\pm$0.03 & 1.08$\pm$0.04 & 0.93$\pm$0.06 & 1.90$\pm$0.16 & 1.70$\pm$0.18 & 0.89$\pm$0.09 \\
26.6--65.3 & 3.47$\pm$0.01 & 0.584$\pm$0.001 & 1.33$\pm$0.01 & 1.07$\pm$0.03 & 0.57$\pm$0.03 & 1.60$\pm$0.05 & 0.94$\pm$0.12 & 0.48$\pm$0.15 & 0.36$\pm$0.11 \\
\hline
\hline\multicolumn{10}{c}{\textbf{\textit{Chandra} Deprojected Narrow-band}} \\
\hline\multicolumn{10}{c}{Deprojection Method: \textit{dsdeproj}}     \\
\multicolumn{10}{c}{Model: \textit{vgadem}}     \\
\hline
0--4.4 & -- & -- & 0.87$\pm$0.13 & 1.09$\pm$0.02 & 0.93$\pm$0.02 & -- & 1.33$\pm$0.18 & 3.30$\pm$0.33 & 3.07$\pm$0.31 \\
4.4--11.0 & -- & -- & 1.45$\pm$0.07 & 1.77$\pm$0.01 & 1.26$\pm$0.01 & -- & 2.49$\pm$0.09 & 2.94$\pm$0.12 & 1.93$\pm$0.08 \\
11.0--26.6 & -- & -- & 1.94$\pm$0.03 & 1.95$\pm$0.01 & 1.30$\pm$0.01 & -- & 2.02$\pm$0.06 & 2.25$\pm$0.09 & 1.17$\pm$0.04 \\
26.6--65.3 & -- & -- & 1.12$\pm$0.01 & 1.15$\pm$0.01 & 0.72$\pm$0.01 & -- & 1.13$\pm$0.04 & 1.31$\pm$0.06 & 1.09$\pm$0.05 \\
\hline
\hline
\end{tabular}
\end{table*}

\section{Discussion}
\label{sec:discussion}

\subsection{Systematic uncertainties and limitations of the analysis}
\label{sec:systematics_n_limitations}
As demonstrated throughout this paper, a careful analysis of the spectra is extremely important for accurately measuring metal abundances and their distribution. Indeed, if the spectra are fitted using overly simplistic assumptions, the subsequent incorrect determination of the metal abundance profiles can easily produce artificial trends or, on the contrary, wipe out actual features. From our analyses, we report that the following factors can significantly affect the measurements of the metal abundance drops:

\begin{itemize}
    \item Temperature distribution effects: An incorrect modeling of the thermal structure of the ICM can  simultaneously affect the continuum emission and the modeling of the mostly unresolved spectral lines. This naturally results in incorrect abundance determinations. One of the best examples is the Fe bias which leads to an underestimated Fe abundance when using a single-temperature model to describe a multi-temperature plasma \citep[][see also Mernier et al. \citeyear{Mernier2018}]{Buote2000}. As shown by \citet{Werner2006}, this Fe bias can easily create artificial Fe drops in some cases. In addition, our present analysis shows that even the choice of the multi-temperature model can alter the shape of the abundance drops (see e.g., the Fe drop in Fig.~\ref{fig:deproj_vgadem_vs_2Tvapec}). This effect is nevertheless less important than the Fe bias.
    
    \item Projection effects: Projection along the line of sight, especially in systems with strong temperature gradients, can lead to the mixing of the emission of the cooler gas in the cluster center with the emission of the hotter ICM at larger radii. The projected view of the central low-metallicity gas overlapping with the higher-metallicity outer layers will essentially result in overestimation of the abundances in the central regions. These projection effects can explain the relatively weaker central drops seen in the projected as compared to the deprojected profiles (Fig.~\ref{fig:deproj_vgadem_vs_2Tvapec}). As shown in Fig.~\ref{fig:XMM_vs_Chandra_deproj_projct_vs_dsdeproj}, different deprojection techniques result in profiles with consistent shapes, but may add some extra uncertainties to the measurements of the absolute abundance values. 
    
    \item Continuum effects: We found that the abundances of weak lines can be significantly over-/underestimated, even by a slightly incorrect modeling of the local continuum level (see Sect. \ref{sec:narrow_band_analys} and Fig.~\ref{fig:narrow_vs_broadband}). As explained in Sect. \ref{sec:narrow_band_analys}, the best way to deal with this effect is to fit these lines individually, within narrow energy bands. In the case of Ar, this effect is significant and clearly affects the interpretations and conclusions of the present study (see also Sect. \ref{sec:origin_of_metal_drop}). Therefore, we caution that narrow-band fits should be systematically considered when one aims to measure accurate abundances from very deep observations of X-ray bright clusters. 
    
    In our case, the Fe abundances measured with only the K-shell transitions are formally consistent with those measured using the full energy band. We note, however, that in the central cooler regions of the ICM, Fe abundances are often derived from the Fe-L complex, which is very sensitive to the modeling uncertainties of the temperature structure of the plasma. A possible mismatch between broad-band and narrow-band derived Fe abundances in other systems may thus reveal a more complicated temperature structure than initially assumed.
    
    \item Instrumental calibration, data quality, and detector PSF: Because no X-ray mission is perfectly calibrated, comparing measurements done with the same analysis methods using two or more instruments is strongly recommended for very deep data. For instance, it is well known that \textit{XMM-Newton} and \textit{Chandra} do not measure the same X-ray temperatures \citep{Schellenberger2015}. In the Centaurus cluster, we show that these two telescopes sometimes provide slight but significant differences in metal abundances as well. Finally, although these effects should be limited in our analysis, we note that the spatial resolution of the instrument can also play an important role as a broader PSF can smooth out gradients.
\end{itemize}

Considering the above factors, we argue that the deprojected narrow-band analysis presented in Sect. \ref{sec:narrow_band_analys}, fitted with the Gaussian-shaped multi-temperature model \textit{vgadem}, and based on the \textit{Chandra} data (i.e., taking advantage of its unprecedented PSF of $\sim$0.5"), provides the most reliable measurements to date of the central abundance drops -- especially when emission lines are weak. We note however, that some limitations still remain and could be resolved only with future missions and instruments:
\begin{itemize}
    \item As the ICM becomes more multi-temperature when approaching the cluster center, it is not clear how the metal abundances should be associated with the different thermal components. With the current spectral resolution of CCDs, assuming different abundances for two or more individual (single-temperature) components would considerably increase the degeneracy between parameters. Determining how complex abundance distributions would affect the central drops requires comprehensive high-resolution simulations that need to be analyzed using mock spectra \citep[for an example of simulated clusters investigated via mock \textit{XMM-Newton} data, see e.g.,][]{Rasia2008}. 
    
    \item While in our analysis we have tried two simplified assumptions to model the multi-temperature plasma (i.e., the 2T \textit{vapec} and the \textit{vgadem} models), the actual emission might be different from both these assumptions, and might even have an asymmetric emission measure distribution. Determining the real temperature distribution in the ICM is far from trivial -- even when the spectral resolution is considerably improved \citep{Hitomi2018} -- and probably requires a very good spatial resolution as well.
    
    \item All deprojection methods rely on the assumption of spherical symmetry. Based on the relatively regular morphology of the X-ray emission of the Centaurus cluster, this seems to be a reasonable assumption and the observed modest departures from spherical symmetry are unlikely to affect our conclusions.  
\end{itemize}

More generally, the careful successive checks performed in the present study have shown that the presence of drops in metal abundance is strongly established. It is also important to note that, since for these very deep data sets the systematic uncertainties dominate over the statistical ones, adding more \textit{Chandra} and/or \textit{XMM-Newton} exposures in the future is not expected to significantly improve the accuracy of the present measurements. While observations with future instruments will be absolutely necessary for this purpose, we speculate that all the limitations discussed above would somewhat affect the magnitude of the metal abundance drops, but are not expected to significantly change  our conclusions.

\subsection{The origin of the metal abundance drops}
\label{sec:origin_of_metal_drop}

Using the best data that are available to date for the Centaurus cluster, our comprehensive analysis confirms the central drops seen in the abundances of the reactive elements Fe, Si, S, Mg and Ca, with a much greater significance than obtained previously  \citep[see][]{Sanders2002,Sanders2006b,Panagoulia2013,Sanders2016}. After adopting our preferred analysis approach (i.e., \textit{vgadem} multi-temperature model, \textit{ACIS-S} spectra deprojected using \textit{dsdeproj}, and abundances fitted individually within a narrow-band), we have checked carefully that these drops do not disappear when assuming a departure from these assumptions. Using narrow-band fits, the Fe, Si, S and Mg drops are seen inside a region of $\sim$10 kpc from the center. For these elements, the spatial extent of the drops is broadly consistent with \citet{Panagoulia2013}. It is also interesting to note that the inner region, where the bulk of the central metals is missing, overlaps remarkably well with the dust emission region seen by \citet{Mittal2011} (see also Fig.~\ref{fig:centaurus_rgb}). However, the Ar abundance is fully consistent with {no} drop at the center. Instead, the deprojected narrow-band fits (Fig.~\ref{fig:narrow_vs_broadband}) show a monotonic centrally increasing trend for Ar. More importantly, even taking all the above systematic uncertainties into account, we always see a significant increase in the Ar/Fe ratio towards the center (at least $>2\sigma$ from $\sim$60 kpc to $\sim$2 kpc).
    
As discussed in Sect. \ref{sec:intro}, the radial distribution of Ar across the location of the Fe drop constitutes a crucial prediction of the dust-depletion model in the central ICM, as proposed by \citet{Panagoulia2013,Panagoulia2015}. Together with the observed significant central drops in the abundances of reactive elements Fe, Si, S, Mg and Ca, the peaked profile seen in the central region for a noble gas like Ar, and the central increase of the Ar/Fe abundance ratio, strongly suggest that a significant fraction of the cooled reactive metals is indeed getting deposited into the dust grains present in the central regions of the Centaurus cluster. 

Qualitatively, our observations seem thus to indicate that metal abundance drops are real and form via the above channel. As explained in Sect. \ref{sec:intro}, the drops may also become more pronounced due to mechanical feedback from the central AGN which can uplift the metal-rich dust grains out of the central regions towards outer hotter regions, where they get re-heated and start emitting in the X-ray band again. 

Another open question is the mismatch in the spatial extent of the abundance drops (e.g., Fig.~\ref{fig:narrow_vs_broadband}) and the spatial extent of the dust emission observed in the far-infrared (Fig.~\ref{fig:centaurus_rgb}). While the metal abundance drops typically extend out to $\sim$10 kpc \citep[possibly somewhat further; e.g.,][]{Panagoulia2013}, the dust emission is only seen within $\sim$4 kpc, that is, it is entirely contained within our innermost spectral extraction region. It is possible, however, that a significant amount of dust resides outside $\sim$4 kpc, where its associated infrared emission is too weak to be firmly detected with \textit{Herschel} and \textit{Spitzer}. Another possibility is that most of the dust uplifted by the AGN activity to radii between $\sim$10 kpc and $\sim$27 kpc is destroyed by the interaction with the ICM. In that case, the metal abundance peak, observed in the third shell, is due to the excess of metals that have freshly returned to the ICM.

Finally, we note that the presence of a central Ar drop would not necessarily rule out the existence of a depletion of the central ICM-phase metals and their incorporation into dust grains. As long as the Ar/Fe abundance ratio is constantly increasing towards the cluster core, dust could play a significant role in lowering the central ICM abundances.

\section{Summary}
\label{sec:conclusion}

In this study, we performed detailed, projected and deprojected spectral analyses of very deep \textit{Chandra} and \textit{XMM-Newton} X-ray observations of the Centaurus cluster. We also investigated the various factors that could bias the metal abundance measurements. These factors include an incorrect modeling of the temperature distribution of the plasma, projection along the line of sight, uncertainties in the modeling of the local continuum (especially when weak emission lines are fitted over a broad energy band), poor data quality, as well as detector PSF and calibration. 

When adopting our most conservative approach, taking most of these systematic effects into account, we confirm the central drops seen in the abundance profiles of the reactive elements such as, Fe, Si, S, Mg and Ca, with a $>6\sigma$ confidence. For Fe, Si, S, and Mg, the drops are seen at $r\sim10$ kpc, while for Ca, a significant drop is seen only at $r\sim4$ kpc. The region of these metal abundance drops therefore broadly coincides with the central dust emission. Additionally, we find a centrally increasing/flattening trend in the abundance of the noble gas Ar, further confirmed by a significant increase in the Ar/Fe ratio towards the innermost region of the cluster. These results constitute strong observational support in favor of the scenario proposed by \citet{Panagoulia2013,Panagoulia2015} in which the drops in central metal abundance are due to the incorporation of a significant fraction of the cooling reactive elements into dust grains, leading to their disappearance from the X-ray band.

\section*{Acknowledgements}

This work was supported by the Lend\"ulet LP2016-11 grant awarded by the Hungarian Academy of Sciences. SRON is supported financially by NWO, the Netherlands Organization for Scientific Research.  This research has made use of software provided 
by the \textit{Chandra X-ray Centre} (CXC) in the application packages CIAO, CHIPS and Sherpa. 
The scientific results reported in this article are based to a significant degree on data, software and web tools obtained from the High Energy Astrophysics 
Science Archive Research Center (HEASARC), a service of the Astrophysics Science Division at 
NASA/GSFC and of the Smithsonian Astrophysical Observatory's High Energy Astrophysics Division. The scientific results reported in this article are based on observations made by the {\it Chandra} X-ray Observatory. This work is based on observations obtained with \textit{XMM-Newton}, an ESA science mission with instruments and contributions directly funded by ESA member states and the USA (NASA).

\bibliography{CCMD_final.bbl
}

\end{document}